\documentclass{aa}  
\usepackage{comment}
\usepackage{graphicx}
\usepackage{color}
\usepackage{txfonts}
\usepackage{natbib}
\usepackage[squaren,Gray]{SIunits}
\usepackage{ulem}

\begin{document} 
  \renewcommand{\vec}[1]{\mathbf{#1}}
  \newcommand{\diver}{\vec{\nabla} \cdot}  
  \newcommand{\divs}{\vec{\nabla}}  
  \newcommand{\grad}{\vec{\nabla}}  
  \newcommand{\vgrad}{(\vec{v} \cdot \vec{\nabla})}  
  \newcommand{\dvgrad}{(\vec{\Delta v} \cdot \vec{\nabla})}  
  \newcommand{\dvkgrad}{(\vec{\Delta v}_k \cdot \vec{\nabla})}  
  \newcommand{\vggrad}{(\vec{v_{\mathrm{g}}} \cdot \vec{\nabla})}  
  \newcommand{\vdgrad}{(\vec{v_{\mathrm{d}}} \cdot \vec{\nabla})}  
\newcommand*\Laplace{\mathop{}\!\mathbin\bigtriangleup}
\newcommand\nature{Nature}

\def\red#1{{\color{red} #1}}
\def\blue#1{{\color{blue} #1}}
   \title{Protostellar collapse: the conditions to form dust rich protoplanetary disks. }
\titlerunning{Protostellar collapse: conditions for dust rich protoplanetary disks. }

    \author{U. Lebreuilly
          \inst{1}
          \and
          B. Commer\c con\inst{1}
          \and G. Laibe \inst{1}}

   \institute{\'Ecole normale sup\'erieure de Lyon, CRAL, UMR CNRS 5574, Universit\'e de Lyon , 46 All\'ee d Italie, 69364 Lyon Cedex 07, France
              \\
              \email{ugo.lebreuilly@ens-lyon.fr}      
             }

   \date{}


  \abstract
    {Dust plays a key role during star, disk and planet formation. Yet, its dynamics during the protostellar collapse remains a poorly investigated field. Recent studies seem to indicate that dust may decouple efficiently from the gas during these early stages.}
   {We aim to understand how much and in which regions dust grains concentrate during the early phases of the protostellar collapse, and see how it depends on the properties of the initial cloud and of the solid particles.}
   {We use the multiple species dust dynamics solver \textsc{multigrain} of the grid-based code \texttt{RAMSES} to perform various  simulations of dusty collapses. We perform hydrodynamical and magnetohydrodynamical simulations where we vary the maximum size of the dust distribution, the thermal-to-gravitational energy ratio and the magnetic properties of the cloud. We simulate the simultaneous evolution of ten neutral dust grains species with grain sizes varying from a few nanometers to a few hundredth of microns.}
   {We obtain a significant decoupling between the gas and the dust for grains of typical sizes a few $\sim 10~\micro\meter$. This decoupling strongly depends on the thermal-to-gravitational energy ratio, the grain sizes or the inclusion of a magnetic field. With a semi-analytic model calibrated on our results, we show that the dust ratio mostly varies exponentially with the initial Stokes number at a rate that depends on the local cloud properties.}
  {We find that larger grains tend to settle and drift efficiently in the first-core and in the newly formed disk. This can produce dust-to-gas ratios of several times the initial value. Dust concentrates in high density regions (cores, disk and pseudo-disk) and is depleted in low density regions (envelope and outflows). The size at which grains decouple from the gas depends on the initial properties of the clouds. Since dust can not necessarily be used as a proxy for gas during the collapse, we emphasize on the necessity of including the treatment of its dynamics in protostellar collapse simulations.}
  

 \maketitle
%

\section{Introduction}

Small dust grains are essential ingredients of star, disk and planet formation. They regulate the thermal budget of star forming regions through their opacity and thermal emission  \citep{2007ARA&A..45..565M,2004tcu..conf..213D}. In addition, they are thought to be the main formation site of H$_2$ at present days \citep{1963ApJ...138..393G}. It is widely accepted that planet formation is induced by dust growth within protoplanetary disks \citep[see the recent review by ][]{2016SSRv..205...41B}.  Finally, the dust grains are significant charge carriers \citep{2016A&A...592A..18M,2016MNRAS.457.1037W,2016MNRAS.460.2050Z}  and therefore regulate the evolution of magnetic fields during the protostellar collapse which can affect, among others, the disk formation \citep{2016A&A...587A..32M,2020A&A...635A..67H} and the fragmentation process \citep{2011ApJ...742L...9C}.

Until recently, one paradigm was that dust of the interstellar medium (ISM) is usually composed of grains with sizes up to $\sim 0.1 ~\micro\meter$ with a typical size distribution well modelled by the Mathis-Rumpl-Nordsieck distribution \citep[MRN,][]{1977ApJ...217..425M}. Recent observations seem to  indicate that larger grains could exist in the denser regions of the ISM. \cite{2010Sci...329.1622P} proposed that over-bright envelopes of prestellar cores (coreshine) could be explained by the presence of micrometer grains. In addition, it was suggested that recent observations with ALMA of the polarised light at (sub)millimeter wavelengths in Class 0 and I objects could be interpreted as the presence of grains up to $\approx 100~ \micro\meter$ \citep{2015ApJ...809...78K,2016ApJ...831L..12K,2016A&A...593A..12P,2018ApJ...859..165S,2018ApJ...869..115S,2019ApJS..245....2S,2019MNRAS.488.4897V}.  \cite{2019arXiv191004652G} has also shown that the low values of the dust emissivity in Class 0 objects could indicate the presence of these large grains in their envelope. Finally, \cite{2020arXiv200602812T} estimated that the mass of solids in Class 0 disks is sufficient to grow planets only if large grains are included in the opacity models, which might indicate dust growth in the early phases of protostar formation. 

 Over the past few years, significant improvements have been made in numerical models to better understand the early phases of the protostellar collapse that leads to the first Larson core formation \citep{1969MNRAS.145..271L}. The angular momentum budget is a long-standing problem in star formation. Indeed, the specific angular momentum of prestellar cores differs to those of young stars by more than three orders of magnitude \citep{1995ARA&A..33..199B,2013EAS....62...25B}. In numerous studies, the magnetic braking has been investigated as one of the possible solutions to address this issue \citep{2003ApJ...599..351A,2007Ap&SS.311...75P,2008A&A...477....9H,2011ApJ...742L...9C,2016A&A...587A..32M}. State-of-the-art simulations account now for the effect of magnetic fields both in a ideal \citep{2010A&A...510L...3C} and non-ideal \citep{2015ApJ...801..117T,2018arXiv180108193V,2019MNRAS.489.1719W} magnetohydrodynamics  (MHD) framework, radiative feedback \citep{2010A&A...510L...3C,2015A&A...578A..12G,2015ApJ...801..117T} and other physical mechanisms. Only \citet{2017MNRAS.465.1089B} have investigated the dynamics of dust during the 3D protostellar collapse (\textsc{dustycollapse}). They report that $\sim 100~\micro\meter$ grains can significantly decouple from the gas leading to large increase of dust-to-gas ratio in the disk and the first Larson core.  In 2D, \cite{2019A&A...631A...1V} have studied the gas and dust decoupling in gravitoviscous protoplanetry disks including dust growth and also report strong variations of dust-to-gas ratio. So far, no 3D \textsc{dustycollapse} simulation has been performed in a MHD/non ideal MHD context or with multiple dust species. 

Multifluid-multi-dust species simulations are prohibitive as they require to solve two additional equations per dust species (mass and momentum conservation). Recent theoretical \citep{2014MNRAS.440.2136L,2014MNRAS.440.2147L} and numerical \citep{2018MNRAS.476.2186H} developments have shown that, for strongly coupled dust and gas mixtures, only the mass conservation equation needs to be solved as the differential velocity between gas and dust can be directly determined by the force budget on both phases. This is the so-called one-fluid or monofluid approach in the diffusion limit \citep{2014MNRAS.440.2136L}. In a recent study \citep{2019A&A...626A..96L}, we presented a fast, accurate and robust implementation of dust dynamics for strongly coupled gas and dust mixtures that allows an efficient treatment of multiple grain species in the adaptive-mesh-refinement \citep[AMR,][]{1984JCoPh..53..484B} finite-volume code \texttt{RAMSES} \citep{2002A&A...385..337T}. In this  study, we extend this one-fluid formalism to neutral dust grains in a partially ionized plasma. We present the first simulations of protostellar collapse of gas and dust mixtures with multiple grain species (\textsc{multigrain} hereafter). In particular, we investigate how the maximum grain size of the dust distribution, the ratio between the thermal and gravitational energy, i.e. the thermal support,  and the magnetic fields may affect the decoupling between the gas and neutral dust grains.

This paper is organised as follows. We recall the general framework of gas and dust mixtures dynamics and extend it to neutral grains in the presence of a magnetic field in Sect. \ref{sec:Fr} and the methods in Sect. \ref{sec:MET}. Section \ref{sec:RES} gives an overview of the different models considered. We summarize the important features of the dusty-collapses obtained in our simulations in Sect. \ref{sec:FEAT}. We show that dust grains have very distinct behaviors in the core and the fragments, the disk and the pseudo-disk, the outflow and the envelope. In Sect. \ref{sec:ana}, we propose a semi-analytical and a simplified analytical model to estimate the central dust enrichment during the collapse. We discuss the implications and caveats of this work in Sect. \ref{sec:DISC}. Finally we present our conclusions and perspectives in Sect. \ref{sec:CONCLU}.

\section{Framework}
\label{sec:Fr}

\subsection{Dusty hydrodynamics for the protostellar collapse}
\label{sec:theory}

A gas and dust mixture with $\mathcal{N}$ small grains species can be modeled as a monofluid in the diffusion approximation \citep{2014MNRAS.444.1940L,2015MNRAS.454.2320P,2018MNRAS.476.2186H,2019A&A...626A..96L}. This fluid, of density $\rho$, flows at its barycenter velocity $\vec{v}$. The $k$-th dust phase, of density $\rho_{k}$, has the specific velocity $\vec{v}+\vec{w_k}$, where $\vec{w_k}$ is the differential velocity between the dust and the barycenter. In the context of the protostellar collapse and in the absence of magnetic fields, this mixture is well described by the following set of equations
\begin{eqnarray} 
  \label{eq:simpl_cons}
  \frac{\partial \rho}{\partial t}  + \diver \left[ \rho \vec v \right]&=&0, \nonumber  \\   \frac{\partial   \rho_{k} }{\partial t} +\diver \left[ \rho_{k} \left( \vec{v} +  \vec{w_k}\right) \right] &=& 0, \ \forall k \in \left[1,\mathcal{N}\right], \nonumber \\
  \frac{\partial \rho \vec{v}}{\partial t} + \diver \left[P_{\mathrm{g}} \mathbb{I} + \rho (\vec{v}\otimes  \vec{v}) \right] &=&-\rho \nabla \phi,
 \end{eqnarray}
 where $P_{\mathrm{g}}$ is the thermal pressure of the gas and $\mathbb{I}$ the identity matrix. The gravitational potential $\phi$ is set by the Poisson equation
 \begin{eqnarray}
  \Laplace{\phi}= 4 \pi \mathcal{G} \rho,
 \end{eqnarray}
where $\mathcal{G}$ denotes the gravitational constant.

The previous equations are closed using a barotropic law that reproduces both the isothermal regime at low density and the adiabatic regime when the density reaches the critical value $\rho_{\mathrm{ad}}$ which corresponds to the density at which dust becomes opaque to its own radiation \citep{1969MNRAS.145..271L}. Similarly to \cite{2008A&A...482..371C}, we express the gas pressure as
 \begin{eqnarray}
P_{\mathrm{g}}=\rho_{\mathrm{g}} c_{\mathrm{s,iso}}^2  \left[1+ \left(\frac{ \rho_{\mathrm{g}} }{\rho_{\mathrm{ad}}} \right)^{\gamma-1}\right],
 \end{eqnarray}
The gas density $\rho_{\mathrm{g}}$ is $\rho_{\mathrm{g}}=\rho-\sum_k  \rho_{k}$. Regions of low densities are isothermal and have for sound speed $c_{\mathrm{s,iso}}$.

 As in \cite{2019A&A...626A..96L}, we model a single dust grain $k$ as a small compact and homogeneous sphere of radius $s_{\mathrm{grain},k}$ and intrinsic density $\rho_{\mathrm{grain},k}$ \footnote{distinct from the dust density $\rho_k$}. When the grain is smaller than the mean free path of the gas (the so-called Epstein drag regime, \citealt{1924PhRv...23..710E}), the drag stopping time $t_{\mathrm{s},k}$ is given by 
   \begin{eqnarray}
   \label{eq:tstop}
   t_{\mathrm{s},k} \equiv \sqrt{\frac{\pi \gamma}{8}} \frac{\rho_{\mathrm{grain},k}}{\rho}\frac{s_{\mathrm{grain},k}}{c_{\mathrm{s}}} ,
 \end{eqnarray}
 where $\rho$ is the total density of the gas and dust mixture, $c_{\mathrm{s}}$ is the sound speed of the gas and $\gamma$ its adiabatic index. 
 
 If the differential velocity $\Delta \vec{v}_{\mathrm{k}} \equiv \vec{v}_{\mathrm{k}}- \vec{v_{\mathrm{g}}} $ between the gas and the dust is supersonic, a correction in the drag regime must be applied. In this case the stopping time is given by \citep{1975ApJ...198..583K}

\begin{eqnarray}
     \label{eq:tstopkw}
   t_{\mathrm{s},k} \equiv \sqrt{\frac{\pi \gamma}{8}} \frac{\rho_{\mathrm{grain},k}}{\rho}\frac{s_{\mathrm{grain},k}}{c_{\mathrm{s}}} \left(1 + \frac{9}{128 \pi} {\mathcal{M}_{\mathrm{d}}}^2 \right)^{-1/2} ,
\end{eqnarray}
where  $\mathcal{M}_{\mathrm{d}} = \frac{|\Delta \vec{v}_{\mathrm{k}}|}{c_{\mathrm{s}}} $ is the differential velocity Mach number. In the remaining of this paper, unless specified, we consider this correction.

In the terminal velocity approximation, the differential velocity of the phase $k$ is
\begin{eqnarray}
   \label{eq:wk}
\vec{w_k} = \left[\frac{\rho}{\rho-\rho_k}t_{\mathrm{s},k}- \sum_{l=1}^{\mathcal{N}} \frac{\rho_l}{\rho-\rho_l} t_{\mathrm{s},l}\right] \frac{\nabla P_{\mathrm{g}}}{\rho},
 \end{eqnarray}
and the gas and dust velocities, $\vec{v_{\mathrm{g}}}$ and $\vec{v_{k}}$ are given by 
\begin{eqnarray}
 \vec{v_{\mathrm{g}}} &=& \vec{v}- \sum_k \frac{\rho_{k}}{\rho-\rho_{k}}\vec{w_k},\nonumber \\
 \vec{v_{k}} &=& \vec{v}+\vec{w_k}.
\end{eqnarray}
For later purposes, we define the dust ratio  $
  \epsilon_k \equiv \frac{\rho_{k}}{\rho},
$ the total dust ratio 
  $  \epsilon \equiv \sum_k^{\mathcal{N}}\epsilon_k,
 $ and the dust-to-gas ratio
   $ \theta_{\mathrm{d}} \equiv \frac{\sum_k^{\mathcal{N}}\rho_{k}}{\rho_{\mathrm{g}}}$. For any quantity $A$, we define $\bar{A}\equiv A / A_0$ where $A_0$ is its initial value. $\bar{\epsilon}$ and $\bar{\theta}$ are called the dust-ratio and dust-to-gas ratio enrichment respectively.
Further details about the monofluid formalism and its implementation in \texttt{RAMSES} can be found in \citet{2019A&A...626A..96L}.

 \subsection{Dusty-MHD with neutral grains}
 
 We extend the above formalism to neutral grains embedded in a weakly ionized plasma. Here we only consider the resistive effect of ambipolar diffusion, i.e. the drift between ions and neutrals other than the dust. This formalism can be straightforwardly extended to more general Ohm's laws. In this context, the equations of dusty-magnetohydrodynamics with neutral grains (Ndusty-MHD) write
 \begin{eqnarray} 
  \label{eq:simpl_MHD}
  \frac{\partial \rho}{\partial t}  + \diver \left[ \rho \vec{v} \right]&=&0, \nonumber  \\   \frac{\partial   \rho_{k} }{\partial t} +\diver \left[ \rho_{k} \left( \vec{v} +  \vec{w_k}\right) \right] &=& 0, \ \forall k \in \left[1,\mathcal{N}\right], \nonumber \\
  \frac{\partial \rho \vec{v}}{\partial t} + \diver \left[\left(P_{\mathrm{g}}+\frac{\vec{B}^2}{2}\right) \mathbb{I} + \rho (\vec{v}\otimes  \vec{v}) -\vec{B}\otimes \vec{B} \right] &=&-\rho \nabla \phi, \nonumber \\
    \frac{\partial \vec{B}}{\partial t} - \nabla \times \left[(\vec{v}-\sum_k \frac{\rho_{k}}{\rho-\rho_{k}}\vec{w_k})\times \vec{B}\right] & &\nonumber \\
    +\nabla \times \left[ \frac{\eta_{\mathrm{A}} c^2}{4 \pi |\vec{B}|^2}[(\nabla \times \vec{B}) \times \vec{B}] \times \vec{B} \right] &=&0,\nonumber \\
    \nabla \cdot  \vec{B} &=&0,  
 \end{eqnarray}
where $\vec{B}$ is the magnetic field, $\eta_{\mathrm{A}}$ is the ambipolar resistivity and $c$ is the speed of light. We note that $\frac{\eta_{\mathrm{A}} c^2}{4 \pi |\vec{B}|^2}[(\nabla \times \vec{B}) \times \vec{B}]$ is the differential velocity between the ions and the neutrals in the gas phase \citep{2012ApJS..201...24M,2016A&A...587A..32M}. We then must correct the neutrals velocity by accounting for the differential velocity between the gas and the barycenter.  The term $-\sum_k \frac{\rho_{k}}{\rho-\rho_{k}}\vec{w_k}$  appears in the induction equation to take that into account. 

In addition from the pressure force, a magnetic force now applies on the plasma. In a previous work, \cite{2014MNRAS.444.1940L} have given the expression of the differential velocity for a general force budget. Using this formula we find that

\begin{eqnarray}
\label{eq:wkB}
 \vec{w_k} = \left[\frac{\rho}{\rho-\rho_k}t_{\mathrm{s},k}- \sum_{l=1}^{\mathcal{N}} \frac{\rho_l}{\rho-\rho_l} t_{\mathrm{s},l}\right] \frac{\nabla P_{\mathrm{g}} - (\nabla \times \vec{B})\times \vec{B}}{\rho},
 \end{eqnarray}
we note that this expression is very similar to what \citet{2006A&A...452..751F} found for single dust species mixtures.

To further simplify Eqs (\ref{eq:simpl_MHD}), we assume in this paper that the plasma velocity $\vec{v}$ is the barycenter velocity which is valid when $ \epsilon_k || \vec{w_k} || \ll || \vec{w_k} ||\ll |\vec{v}|$. In this case the induction equation writes
 \begin{eqnarray} 
  \label{eq:ind_MHD}
    \frac{\partial \vec{B}}{\partial t} - \nabla \times \left[\vec{v}\times \vec{B}\right] \
    +\nabla \times \left[ \frac{\eta_{\mathrm{A}} c^2}{4 \pi |\vec{B}|^2}[(\nabla \times \vec{B}) \times \vec{B}] \times \vec{B} \right] &=&0.  
 \end{eqnarray}

\section{Method}
\label{sec:MET}

\subsection{RAMSES}
For this work, we take advantage of the {\ttfamily RAMSES} code \citep{2002A&A...385..337T}. This finite-volume Eulerian code solves the hydrodynamics equations using a second-order Godunov method  \citep{godunov} on an adaptive-mesh-refinement grid  \citep{1984JCoPh..53..484B}. With a proper refinement criteria, the AMR grid is a powerful tool to study multi-scale problems such as the protostellar collapse of a dense core. The {\ttfamily RAMSES} code is very proficient for problems that require a treatment of magnetohydrodynamics \citep{2006JCoPh.218...44T,2006A&A...457..371F,2012ApJS..201...24M,Hall-pierre,2019A&A...631A..66M}, radiation hydrodynamics \citep{2011A&A...529A..35C,2013MNRAS.436.2188R,2014A&A...563A..11C,2015MNRAS.449.4380R,2015A&A...578A..12G,2020A&A...635A..42M} or cosmic rays  \citep{2016A&A...585A.138D,2019A&A...631A.121D}. 

We extended the code to the treatment of dust dynamics with multiple species in the diffusion approximation and terminal velocity regime \citep{2019A&A...626A..96L}. The method, based on an operator splitting technique has been extensively presented and tested. It uses a predictor-corrector MUSCL scheme \citep{1974JCoPh..14..361V} and is second-order accurate in space. The solver can be used to simultaneously and efficiently model several dust species (\textsc{multigrain}).

\subsection{Boss and Bodenheimer test}

We perform Boss and Bodenheimer tests \citep{1979ApJ...234..289B} to follow the dynamics of the dust during the first collapse and first core formation. 
The parameters of the setup are the initial mass of the dense core (or prestellar core) $M_0$, the total dust ratio $\epsilon_0$, the temperature of the gas $T_{\mathrm{g}}$ and a mean molecular weight $\mu_{\mathrm{g}}$. The ratio between the thermal and the gravitational energy $\alpha$ is
\begin{eqnarray}
 \alpha &=& \frac{5}{2}\frac{(1-\epsilon_0) R_{0} }{ \mathcal{G} M_{0} }\frac{k_{\mathrm{B}}T_{\mathrm{g}}}{\mu_{\mathrm{g}} m_{\mathrm{H}}},
 \end{eqnarray}
and sets the initial radius of the cloud $R_{0}$ and its density $\rho_{0} $. In addition, we impose an initial solid body rotation around the $z$-axis at the angular velocity $\Omega_{0}$ by setting the ratio between the rotational and the gravitational energy $\beta$ given by
 \begin{eqnarray}
  \beta &=& \frac{1}{3} \frac{R_{0}^3 \Omega_{0}^2}{\mathcal{G} M_{0} }.
 \end{eqnarray}
Eventually,  we apply an initial  azimuthal density perturbation according to
 \begin{eqnarray}
 \rho = \rho_{0} \left[1+A \cos \left(m \theta\right)\right].
 \end{eqnarray}

In this paper, we aim to investigate the impact of magnetic fields on the dynamics of neutral dust grains in two simulations, one with ideal MHD and one with ambipolar diffusion. For these runs, we impose an uniform magnetic field using the mass-to-flux-to-critical-mass-to-flux-ratio

 \begin{eqnarray}
\mu = \left(\frac{M_{0}}{\Phi}\right)/\left(\frac{M}{\Phi}\right)_{\mathrm{c}},
\end{eqnarray}
 the critical mass-to-flux ratio being given by $(M/\Phi)_{\mathrm{c}}= \frac{0.53}{3 \pi} \sqrt{5/\mathcal{G}} $ \citep{1976ApJ...210..326M}. We set an angle $\phi_{\mathrm{mag}}$ between the magnetic fields and the rotation axis to reduce the efficiency of the magnetic braking.

 \subsection{Dust grain size distributions}
\label{sec:distri}
   In  our \textsc{multigrain} simulations, $\mathcal{N}>1$ dust bins are considered. The dust ratio of each bins is set from power-law distributions 
 \begin{eqnarray}
 \frac{\mathrm{d}\epsilon}{\mathrm{d}s}= \frac{\epsilon_0}{\int_{S_{\mathrm{min}}}^{S_{\mathrm{max}}}s^{3-m} \mathrm{d}s} s^{3-m}, 
 \end{eqnarray}
 with $\epsilon_0$ the total initial dust ratio and, $S_{\mathrm{min}}$ and $S_{\mathrm{max}}$ being the minimum and maximum sizes of the grains present in the medium, respectively. For the standard MRN distribution,  $S_{\mathrm{min}}=5$~nm, $S_{\mathrm{max}}=250$~nm and $m=3.5$. 

 The method described in \cite{2018MNRAS.476.2186H} is used to compute the initial dust ratio and typical grain size of each bin.  A logarithmic grid is used to determine the edges $S_{k}$ of the bins

 \begin{eqnarray}
 \log (S_{k})= \log\left(\frac{S_{\mathrm{max}}}{S_{\mathrm{min}}}\right)\frac{k}{\mathcal{N}}+\log\left({S_{\mathrm{min}}}\right). 
 \end{eqnarray}
 The typical grain size of a bin $k$ required to compute the stopping time is
 \begin{eqnarray}
 \label{eq:bins}
 s_{k}= \sqrt{S_{k}S_{k+1}}.
 \end{eqnarray}
We note that $S_{\mathrm{min}}$ and $S_{\mathrm{max}}$ are the edges of the distribution and must not be confused with the minimum and maximum bin size that are averaged quantities.
 The initial dust ratios in each bin are computed according to
 \begin{eqnarray}
 {\epsilon_0}_{k}= \epsilon_0 \left[ \frac{S_{k+1}^{4-m}-S_{k}^{4-m}}{S_{\mathrm{max}}^{4-m}-S_{\mathrm{min}}^{4-m}}\right].
 \end{eqnarray}

 \subsection{Setup}

\subsubsection{Cloud setup}
We impose initial conditions that are typical of the first protostellar collapse  \citep{1969MNRAS.145..271L} with $T_{\mathrm{g}}=10$~K, $\mu_{\mathrm{g}}=2.31$ and a solar mass cloud. We also set $\gamma= 5/3$ since molecular hydrogen behaves as a monoatomic gas at low temperatures \citep{1997MNRAS.291..578W}. As explained in Sect. \ref{sec:theory}, a barotropic law is used to close Eqs. \ref{eq:simpl_cons} with $\rho_{\mathrm{ad}}=10^{-13} $g cm~$^{-3}$ \citep{1969MNRAS.145..271L}.  Finally, we always set  $m=2$ and  $A=0.1$ to favor fragmentation and the formation of two spiral arms.

For the dust, we always consider $10$ bins with grain sizes distributed according to Sect. \ref{sec:distri}. In all the models $S_{\mathrm{min}}=5$~nm, and $m=3.5$ and $S_{\mathrm{max}}$ is specified individually.  In all our models we set $\rho_{\mathrm{grain}}=1~ \gram~\centi\meter^{-3}$. Finally, we impose an uniform initial total dust-to-gas ratio of  $\theta_{\mathrm{d},0}=0.01$ in all the models.

The two magnetic models have been computed with $\mu = 5$ and $\phi_{\rm{mag}}= 40^{\circ}$.  For the non-ideal MHD model the ambipolar resistivity is computed similarly to the case of reference of \cite{2016A&A...592A..18M}.

\subsubsection{Numerical setup}

We use the  \textsc{hlld} Riemmann solver \citep{2005JCoPh.208..315M} for the barycenter part of the conservation equations with a \textsc{minmod} slope limiter \citep{1986AnRFM..18..337R} for both the gas and the dust. The Truelove criterion \citep[at least 4 point per Jeans length,][]{1997ApJ...489L.179T} must be satisfied to avoid artificial clump formation. We therefore enforce a refinement criteria that imposes at least 15 points per local Jeans length. The grid is initialized to the level $\ell_{\mathrm{min}}=5$ and allows refinement up to a maximum level $\ell_{\mathrm{max}}=16$ (which gives a resolution between $32^3$ and  $65536^3$ cells).

\subsubsection{Analysis of the models}
\label{sec:analy}

We consider that the first hydrostatic core (FHSC) is fully formed when the peak density reaches $10^{-11} ~\gram~\centi\meter^{-3}$ for the first time. We denote the corresponding time $t_{\mathrm{core}}$, and use this definition to compare our models at similar evolutionary stages. We present here the different objects that are observed in our model and their definition in this work. 
\begin{itemize}

\item The first hydrostatic core/the fragments are any object of density larger than  $10^{-12.5}~ \gram\,\centi\meter^{-3}$, as in \cite{2019A&A...626A..96L}. The FHSC or $\mathcal{F}_0$ corresponds to the central fragment. $\mathcal{F}_1$ and $\mathcal{F}_2$ are the secondary fragments/FHSC.

\item The disk $\mathcal{D}$ is the region that satisfies \cite{2012A&A...543A.128J} criterion.  For the analysis, we place ourselves in cylindrical coordinates $(r, \phi, z)$. A region is identified as a disk if it is  Keplerian  ($v_{\phi}> f_{\rm{thre}} v_{r}$), in hydrostatic equilibrium ($v_{\phi}> f_{\rm{thre}} v_{z}$), rotationally supported ($\frac{1}{2} \rho v_{\phi}^2 > f_{\rm{thre}} P_{\rm{g}}$) and dense $\rho > 3.9 \times 10^{-15} ~\gram ~\centi\meter^{-3}$. As in \cite{2012A&A...543A.128J} , we choose $f_{\rm{thre}}=2$.
\item The pseudo-disk $\mathcal{P}$ \citep{1993ApJ...417..220G}, for magnetic runs, is defined as the regions with $r<5000$~AU and densities above $3.9 \times 10^{-17} ~\gram \centi\meter^{-3}$ that are not in the disk and fragments. The criterion is similar to \cite{2016ApJ...822...12H}. $5000$~AU is an arbitrary distance that is sufficiently larger than the central objects while being smaller than the initial cloud.
\item Jets/outflows $\mathcal{O}$ correspond to any region with $r<5000$~AU with $ \vec{v}\cdot \frac{\vec{r}}{|\vec{r}|}>0.2~\kilo\meter \sec^{-1}$. The criterion is also similar as \cite{2016ApJ...822...12H}.
\item The envelope $\mathcal{E}$ encompasses the regions with $r<5000$~AU that exclude the fragments, the disk/pseudo-disk and the jets/outflows.

\end{itemize}
We consider two different weights for averaging a quantity $A$ over a volume $\mathcal{V}$ in the computational box. 
Volume averaging is computed according to
 \begin{eqnarray}
 \label{eq:vav}
 \left<A\right>_{v} = \frac{\sum_{i \in \mathcal{V}}  \Delta x_i^3 A_i }{\sum_{i \in \mathcal{V}}  \Delta x_i^3}.
 \end{eqnarray}
Mass averaging is computed according to
 
 \begin{eqnarray}
 \label{eq:mav}
  \left<A\right>_{m} = \frac{\sum_{i \in \mathcal{V}} \rho_i  \Delta x_i^3 A_i }{\sum_{i \in \mathcal{V}} \rho_i  \Delta x_i^3},
 \end{eqnarray}
 where $\rho_i$ and $ \Delta x_i $ are the total density and length of individual cells $i$ in the averaged volume. Volume averages emphasize on regions of large spatial extension, i.e. the envelope, while mass averages emphasize on regions of  high density, i.e. the core+disk and the denser regions of the envelope.

\subsection{Regularization of the differential velocity and dust density}
\label{sec:regul}

The terminal velocity approximation is unrealistic in low density regions or in shocked regions where the pressure is discontinuous \citep{2019MNRAS.488.5290L}. We cap the differential velocities to $w_{\mathrm{cap}}$ in our models to avoid prohibitively small timesteps and unrealistically large variations in the dust ratio in strong shock fronts. We impose $w_{\mathrm{cap}}=1~$km~s$^{-1}$. To verify that this does not impact the results, we ran extra models with  $w_{\mathrm{cap}}=0.1$ km~s$^{-1}$, $w_{\mathrm{cap}}=0.5$ km~s$^{-1}$ and  $w_{\mathrm{cap}}=2$ km~s$^{-1}$. A comparison between these models and our fiducial is given in Appendix \ref{ap:apA0}.

In our models, the drift velocity can in some regions be supersonic. To account for the correction presented in Eq (\ref{eq:tstopkw}), we use the drift velocity estimated at the previous timestep to estimate the differential velocity mach number.

Finally, we set the drift velocity to zero at densities lower than the ones of  the initial cloud, i.e. the background. This is a way to ensure that the regions where the terminal velocity is not valid do not affect significantly the calculation.

\subsection{Validity of the model}
  \begin{figure}[t!]
       \centering
          \includegraphics[width=0.5\textwidth]{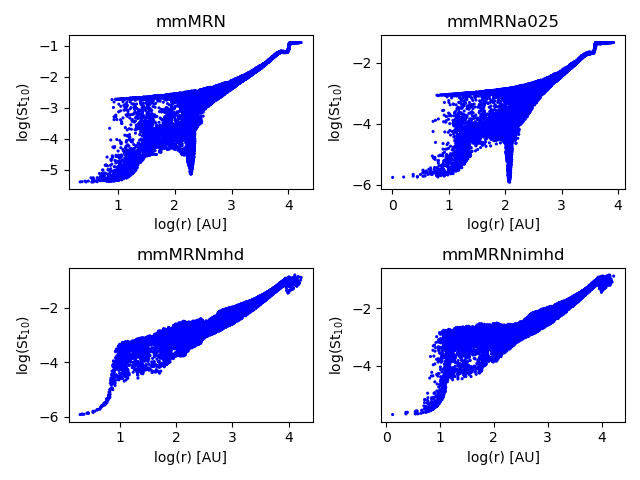}
      \caption{Logarithm of the maximum ($10^{\rm{th}}$ bin) Stokes number  as a function of the spherical radius for the four models with the largest Stokes numbers at $t_{\rm{core}}+2~$kyr. (Top-left) $\textsc{mmMRN}$, (Top-right) $\textsc{mmMRNa0.25}$, (Bottom-left) $\textsc{mmMRNmhd}$ , (Bottom-right) $\textsc{mmMRNnimhd}$. }
      \label{fig:stokes}
      \end{figure}
 
The diffusion approximation is valid as long as the ratio between the stopping time and the dynamical timescale of the gas is small compared to unity \citep{2014MNRAS.444.1940L}. This ratio is called the Stokes number $\mathrm{St}$. 

During the first collapse, the dynamical timescale is the free-fall time $t_{\rm{ff}}$. We have shown in \cite{2019A&A...626A..96L} that for an initial dust-to-gas ratio of $0.01$ and a temperature of $10$~K, the initial Stokes number $\mathrm{St_0}$ of a spherical collapse is given by
\begin{eqnarray}
   \mathrm{St}_0 \sim 0.038\left(\frac{M_0}{1 M_{\odot}}\right) \left(\frac{\rho_{\mathrm{grain}}}{1  \mathrm{\ g } \mathrm{\ cm}^{-3} }\right)  \left(\frac{s_{\mathrm{grain}}}{0.05 \mathrm{\ cm}} \right)  \left({\frac{\alpha}{0.5}}\right)^{3/2}<1.
 \end{eqnarray}
    The dynamics of grains smaller than $0.05 \mathrm{\ cm}$ can be simulated using the diffusion approximation. Note that the Stokes number varies as $\propto \frac{1}{\sqrt{\rho}}$ (since $t_{\rm{ff}} \propto  \frac{1}{\sqrt{\rho}}$ and $t_{\rm{s}}\propto \frac{1}{\rho}$) and hence can increases with a decreasing density. We show in Fig \ref{fig:stokes} the values of the maximum Stokes number as a function of the radius for the four models with the least coupled dust (see Sect. \ref{sec:RES} for a description of the models). The maximum value for St is smaller than $\sim 0.15$ in external regions of the collapse for all our models and it is typically smaller than $ 0.05$ inside the collapsing regions of the models. Rotation provides an additional support to the collapse, which causes an increase of the free-fall timescale. Hence, as the initial angular velocity increases, the initial Stokes number decreases and the diffusion approximation is even more accurate. Similarly, magnetic fields increase the duration of the collapse which broadens the validity domain of the diffusion approximation. Finally, the Stokes number also decreases when $\alpha$ decreases, implying that the diffusion approximation remains valid for small values of $\alpha$. 
     
In the induction equation, we consider for simplicity that the plasma is moving at the barycenter velocity. This approximation is valid when $\epsilon \ll 1$, i.e. when the back-reaction from the dust onto the gas is negligible. Note that we investigate the impact of the back-reaction in Appendix \ref{ap:nback}.

\section{Models}
\label{sec:RES}
The models presented in this section are referenced in Table \ref{tab:models}. All of them have been evolved up to at least $2$~kyr after the formation of the first core. Our fiducial case $\textsc{mmMRN}$ has been run over a longer time.

\begin{table*}[t]
       \caption{Syllabus of the different simulations, with the thermal-to-gravitational energy ratios $\alpha$, maximum grain sizes $S_{\mathrm{max}}$ . The initial mass-to-flux ratio $\mu$ as well as the tilt between the magnetic field and the rotation axis $\phi_{\mathrm{mag}}$ are given for simulations with magnetic field. Additionally, we also provide the formation time of the FHSC $t_{\rm{core}}$ and and the initial Stokes number of the largest grains $\rm{St}_{0,10}$, the mass of the initial core and the number of dust bins. }      
\label{tab:models}      
\centering          
\begin{tabular}{c c c c c c c c }     
\hline\hline       
                   Model & $\alpha$ & $S_{\mathrm{max}}$ (cm) & $\mu $ & $\phi_{\mathrm{mag}}$ ($^{\circ} $) & Ambipolar & $t_{\rm{core}}$ (kyr)  & $\rm{St}_{0,10}$  \\ 
\hline                    
 $\textsc{mmMRN}$  & $0.5$ & $0.1$ & - & - & - & $72.84$ &  $1.22 \times 10^{-2}$ \\
 $\textsc{MRN}$  & $ 0.5 $& $2.5 \times 10^{-5}$ & - & - & - &  $73.6$ & $1.06 \times 10^{-5}$ \\
 $\textsc{100micMRN }$& $0.5$ & $0.01$ &- & - & - & $72.9$ & $1.72 \times 10^{-3}$ \\ 
$\textsc{mmMRNa0.25}$& $0.25$ & $0.1$ & - & - & -& $23$ &   $4.31 \times 10^{-3}$ \\
\hline                    
$\textsc{mmMRNmhd}$  & $0.5$ & $0.1$ & $5$ & $40$ & NO & $81.1$ &    $1.22 \times 10^{-2}$  \\
$\textsc{mmMRNnimhd}$  & $0.5$ & $0.1$ & $5$ & $40$ & YES & $81.1$ &   $1.22 \times 10^{-2}$   \\
\hline \hline
 Initial cloud mass  & Number of dust bins \\ 
 $1 M_{\odot}$ & $10$ \\
 \hline \hline
\end{tabular}
\end{table*}

\subsection{Fiducial simulation}
  \label{sec:fid}
 
 \begin{figure}[t!]
       \centering
          \includegraphics[width=0.5\textwidth]{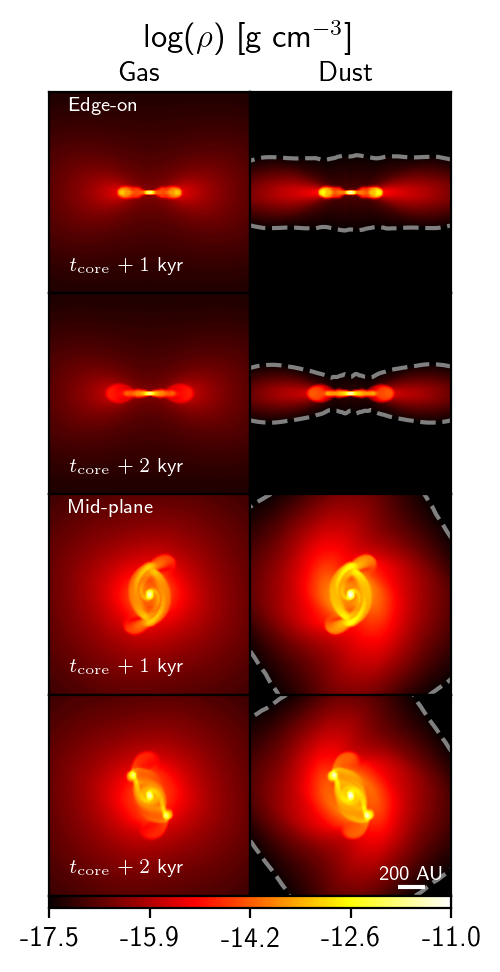}
      \caption{$\textsc{mmMRN}$ test at $t_{\mathrm{core}}+1$~kyr and $t_{\mathrm{core}}+2$~kyr ($t_{\mathrm{core}}= 72.84$~kyr). Edge-on and mid-plane cuts of the gas and the dust densities for the least coupled species are provided (left and right respectively). Values of the gas density are indicated by the colorbar on the bottom. Dust densities have been multiplied by a factor 100 to be represented on the same scale. Hence, colors of the gas and the dust maps match when the dust-to-gas ratio equals its initial value $0.01$. Dust density variations in regions where $\epsilon_0 \rho_{\mathrm{d}}<\mathrm{min}(\rho_{\mathrm{g}})$ have voluntarily not been displayed to highlight the enriched regions. These depleted regions are delimited by the dashed grey lines. This choice of colors applies for all density maps in this study. Gas and dust are clearly not perfectly coupled.}
      \label{fig:fidMRNdens}
      \end{figure}
 
       \begin{figure*}[t!]
       \centering
          \includegraphics[width=\textwidth]{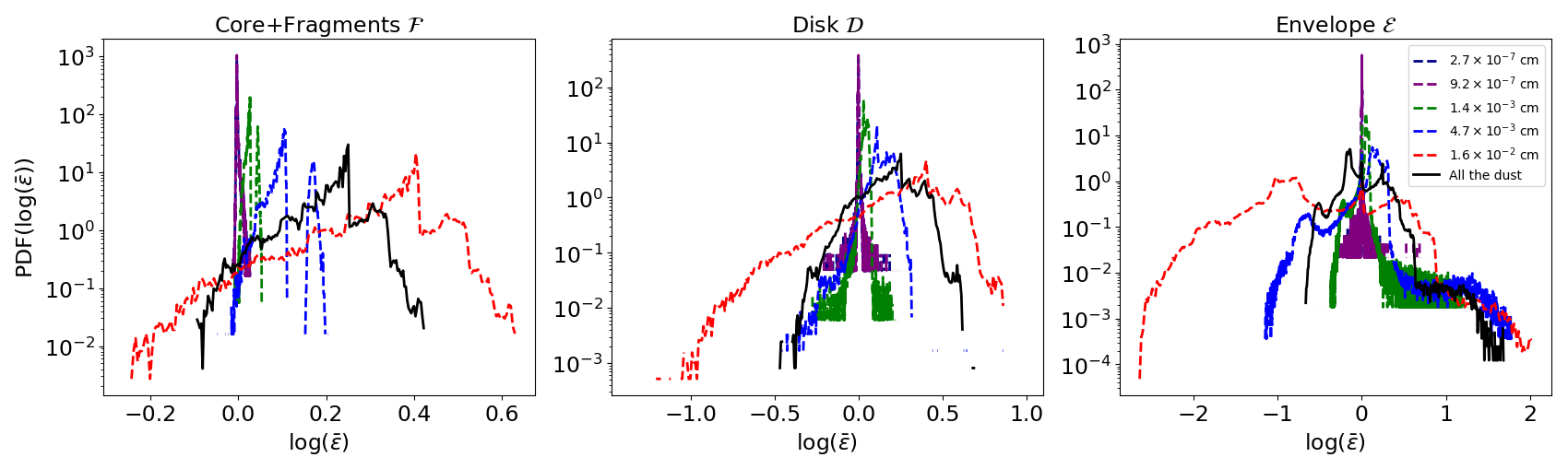}\caption{$\textsc{mmMRN}$ test at $t_{\mathrm{core}}+2$~kyr. Probability density function (PDF) of the dust ratio enrichment $\log(\bar{\epsilon})$  for the two most coupled and three least coupled dust species (colored dashed lines) and for all the dust (black line) in the core+fragments(left), the disk (middle) and the envelope (right). Dust is not a good tracer for the gas here and the dust distribution is not uniform in the considered objects.}
      \label{fig:fidhisteps}
      \end{figure*}    
    
      \begin{figure}[t!]
       \centering
          \includegraphics[width=0.5\textwidth]{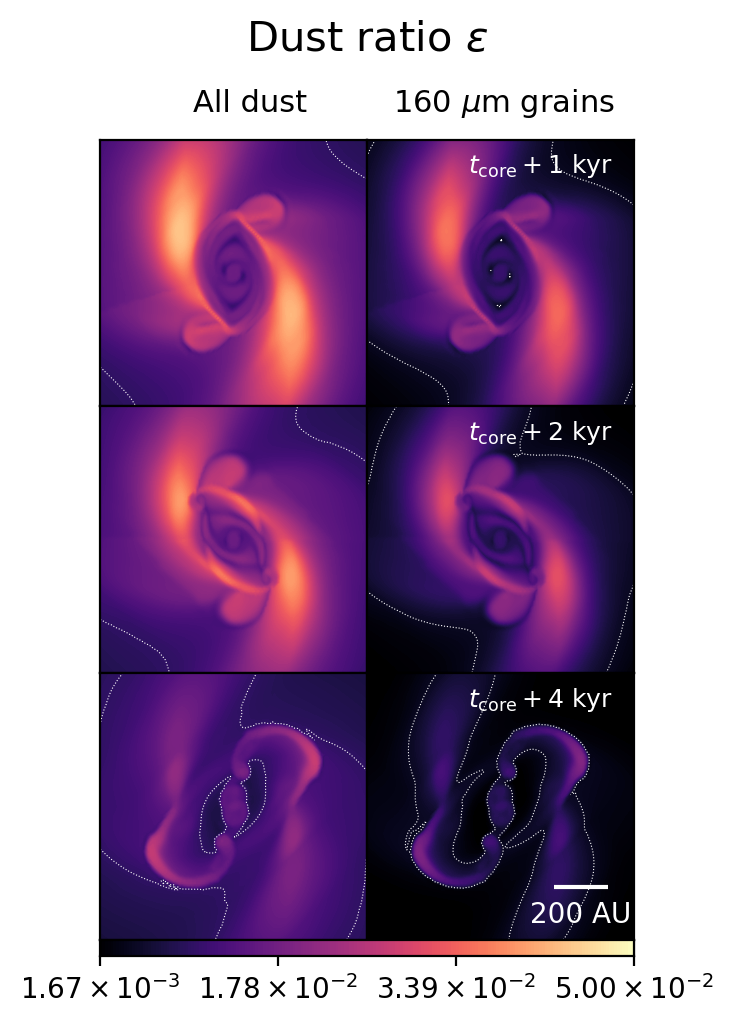}
      \caption{$\textsc{mmMRN}$ test $\sim 1$~kyr (top),  $\sim 2$~kyr (middle) and $\sim 4$~kyr (bottom) after the first core formation ($t_{\mathrm{core}}= 72.84$~kyr). Mid-plane view of the total dust ratio (left) and the dust ratio of the least coupled species (right). The colorbar is the same for both figures. The dotted white lines represent the regions where the total (left) or $160~ \micro\meter $ grains (right) dust-to-gas ratio is at its initial value, which can also be regarded as a dust enrichment line.}
      \label{fig:dratall10}
      \end{figure}

   \begin{figure}[t!]
       \centering
          \includegraphics[width=0.35\textwidth]{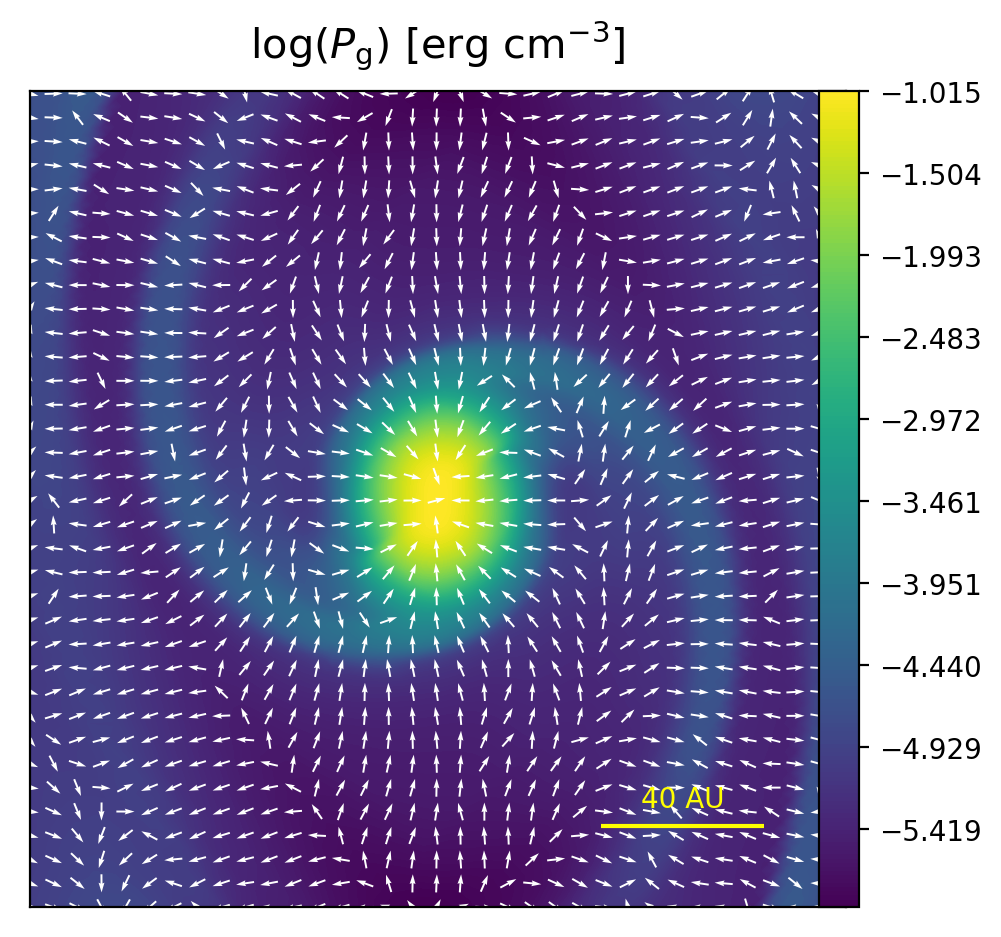}
\caption{$\textsc{mmMRN}$ test at $t_{\mathrm{core}}+1$~kyr ($t_{\mathrm{core}}= 72.84$~kyr). Mid-plane view of the gas pressure (up-close). The white arrows represent the direction of the differential velocity $\vec{w_{10}}$. }
      \label{fig:mmPress}
      \end{figure}

   \begin{figure*}
       \centering
          \includegraphics[width=\textwidth]{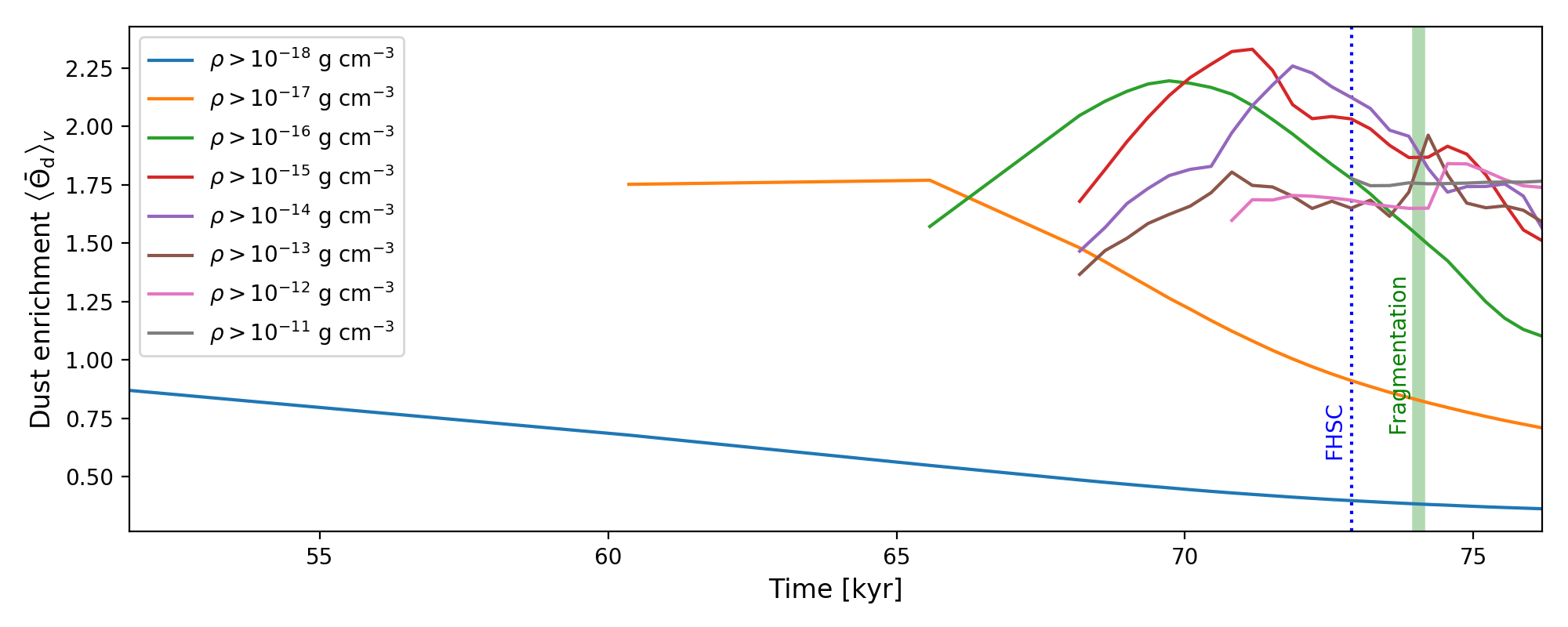}
      \caption{$\textsc{mmMRN}$ test. Volume averaged total dust-to-gas ratio enrichment as a function of time for different density thresholds. The FHSC formation can be identified by the dotted vertical blue line while fragmentation occurs during a time delimited by the green area. We observe a slow decrease of the dust-to-gas ratio at low density at the benefit of an enrichment of high density regions. Cores and fragments at $\rho >10^{-11} ~\gram ~\centi\meter^{-3}$ are formed in a dust rich environment. The dust-to-gas ratio is almost constant for $\rho >10^{-11}~ \gram ~\centi\meter^{-3}$ because the increase of temperature due to the adiabatic contraction strengthen the coupling between the gas and the dust.  }
      \label{fig:enricht}
      \end{figure*}
     
In this section, we present our fiducial case $\textsc{mmMRN}$ where $\alpha=0.5$ and $\beta=0.03$. The grain size distribution is extended up to $S_{\mathrm{max}}=1$~mm. The value of $S_{\mathrm{max}}$ leads to an average size of the last and largest bin $s_{10} \sim 160~\micro\meter$. The total initial dust-to-gas ratio is  $\theta_{\rm{d},0}= 1\%$. 

Figure \ref{fig:fidMRNdens} shows edge-on (four top figures) and mid-plane (four bottom figures) density cuts of the gas (left) and of the dust fluid with the largest grain size ($160~\micro\meter$, right) at $1$~kyr and  $2$~kyr after the formation of the first Larson core ($t_{\mathrm{core}}= 72.84$~kyr for this model), respectively. The density distributions obtained for the gas and the dust are clearly different. This discrepancy originates from an imperfect coupling between the two phases which causes a drift of the dust toward the inner regions of the collapse. This general trend can be explained by a simple force budget on the gas and the dust. Although the gas is partially supported by pressure, dust grains are only subjected to gravity and gas drag. As such, the dust fluid collapses essentially faster than the gas. It therefore enriches the first core and the disk at the cost of a depletion of solids in the envelope. Figure \ref{fig:fidMRNdens} shows that these strong enrichment in the mid-plane and depletion in the envelope have already occurred at $t_{\mathrm{core}}+1$~kyr, and continues for more than $1$~kyr. In the mid-plane, the envelope is enriched in large dust grains close to the central object and depleted further away. In the vertical direction, it is mostly depleted in large grains. In short, after the first core formation these grains are concentrated in a very thin layer of $10-100$~AU above/under the mid-plane. At this stage, the envelope is mostly a reservoir of low dust densities for the large grains. Hence accretion of dust arising from the envelope does not enrich significantly the fragments and the disk in large grains and the dust-to-gas ratio in the disk even decreases.  We note that, the latter is still very enriched by the end of the calculation. Most of the enrichment of dust-to-gas ratio in the central objects is indeed actually taking place during the initial phases of the collapse, when the densities are low everywhere and the coupling between the gas and the dust is the weakest.

 Figure \ref{fig:fidhisteps} shows probability density functions (PDF) of the dust ratio enrichment, denoted $\log(\bar{\epsilon})$. It compares the distributions of the two most coupled and the three least coupled dust species (colored dashed lines) at $t_{\mathrm{core}}+2$~kyr. It also indicates the PDF integrated over the grain size distribution (black line). These PDF are displayed in three different regions, namely the core and the fragments (left), the disk (middle) and the envelope (right). Figure \ref{fig:fidhisteps} shows that the dust enrichment in the inner regions is size-dependent. Small grains experience larger drag that reduces their differential velocity with respect to the gas.
 
Here, grains with sizes smaller than a few microns remain very well coupled with the gas in all the considered regions, whereas for larger grains the PDF of the dust ratio is broad. For $160~ \micro\meter$ grains, the dust ratio increases by one order of magnitude in some regions of the disk and up to two orders of magnitude in the envelope. On average, the dust ratio is $0.018$ in the core, $0.0175$ in the disk and $\sim 0.0086$ in the envelope. In addition, the dust has experienced a strong and local dynamical sorting. We indeed measure a typical standard deviation for the dust-ratio enrichment $\bar{\epsilon}$ of $0.072$ in the core, $0.14$ in the disk and $0.23$ in the envelope. The standard deviation is the largest in the envelope, a region which is depleted in dust in the outer regions and enriched close to the disk and fragments. The disk experiences larger variations of dust-to-gas ratio compared to the core. Indeed, the latter is in adiabatic contraction. This induces high temperatures, which in return causes a strong decrease of the Stokes number. Hence, dust is essentially frozen with the gas in the core and $\vec{v}_{\mathrm{d}}\approx\vec{v}_{\mathrm{g}}$. Figure \ref{fig:dratall10} shows the dust ratio for the total dust distribution (left) and for the $10^{th}$ species (right) in the mid-plane of the collapse, at $1$~kyr (top), $2$~kyr (middle) and $4$~kyr (bottom) after the formation of the first Larson core respectively. The maps on the left and on the right are very similar as most of the evolution of the dust ratio is due to the dynamics of the least coupled species, which represents a large fraction of the dust mass (see also the PDF). The structures in the dust ratio observed in Fig. \ref{fig:dratall10} can be interpreted by looking at the thermal pressure distribution shown in Fig. \ref{fig:mmPress}. Dust grains tend to drift toward local pressure maxima (see Fig.\ref{fig:dratall10}, top panel) where the differential velocity is zeroed (see Eq. \ref{eq:wk}). The essential of the variations of the total dust ratio are due to the largest grains, since they represent most of the dust mass and have the largest drift velocities. Hence, a significant fraction of the dust mass in the inner regions is composed with large grains. Finally, we note that $4$~kyr after the first core formation, the average value for the total dust-to-gas ratio is roughly unchanged but generally increasing (of about $1 \%$) in the core and fragments since their formation. For the disk, we note a decrease of dust ratio of $\sim 30\%$ for the largest grains ($\sim 22 \%$ in total) in the disk since the first core formation. In addition, the total dust-to-gas ratio  continues to diminish in the envelope at this time as settling goes on, with a final average value of $\sim 0.0083$. 
 
Figure \ref{fig:enricht} shows the evolution of the dust-to-gas ratio enrichment averaged in volume for different density threshold in the regions where $R<5000$ AU. The dust depletion at large scales in the envelope at low densities is a relatively slow process, which occurs during the entire collapse. Once regions with larger densities are formed, they usually experience a relatively quick enrichment from the dust content of the low density regions, and then a quick depletion in favor of even denser regions. Figure \ref{fig:enricht} shows that fragmentation occurs in a dust-rich environment. A strong enrichment of the volume where $\rho>10^{-13}~\gram ~\centi\meter^{-3}$ delimited by the brown line indeed  happens exactly when fragments form. This explains why these fragment tend to be more dust-rich than the first hydrostatic core. Interestingly, the dust-to-gas ratio does not vary significantly in the volume where $\rho>10^{-11}~\gram ~\centi\meter^{-3}$ which is in adiabatic contraction. Again, the temperatures in this region are high and therefore the Stokes numbers are very low, which significantly slows down the differential dynamics of the gas and the dust. We note that this volume is already dust enriched by the time of its formation by almost a factor of two.

\subsection{Parameter exploration}
\subsubsection{Maximum grain size}
      \begin{figure}[t]
       \centering
          \includegraphics[width=0.5\textwidth]{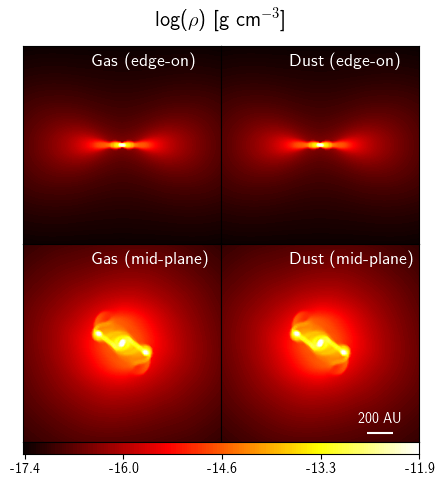}
      \caption{$\textsc{MRN}$ test at $t_{\mathrm{core}}+2$~kyr ($t_{\mathrm{core}}= 73.6$~kyr). Edge-on (top) and mid-plane (bottom) cuts of the gas density (left) and  dust density of the least coupled species (right). The two maps are almost indistinguishable due to the very strong coupling between gas and all dust species.}
      \label{fig:MRNdens}
      \end{figure}
      
      \begin{figure}[t]
       \centering
          \includegraphics[width=0.5\textwidth]{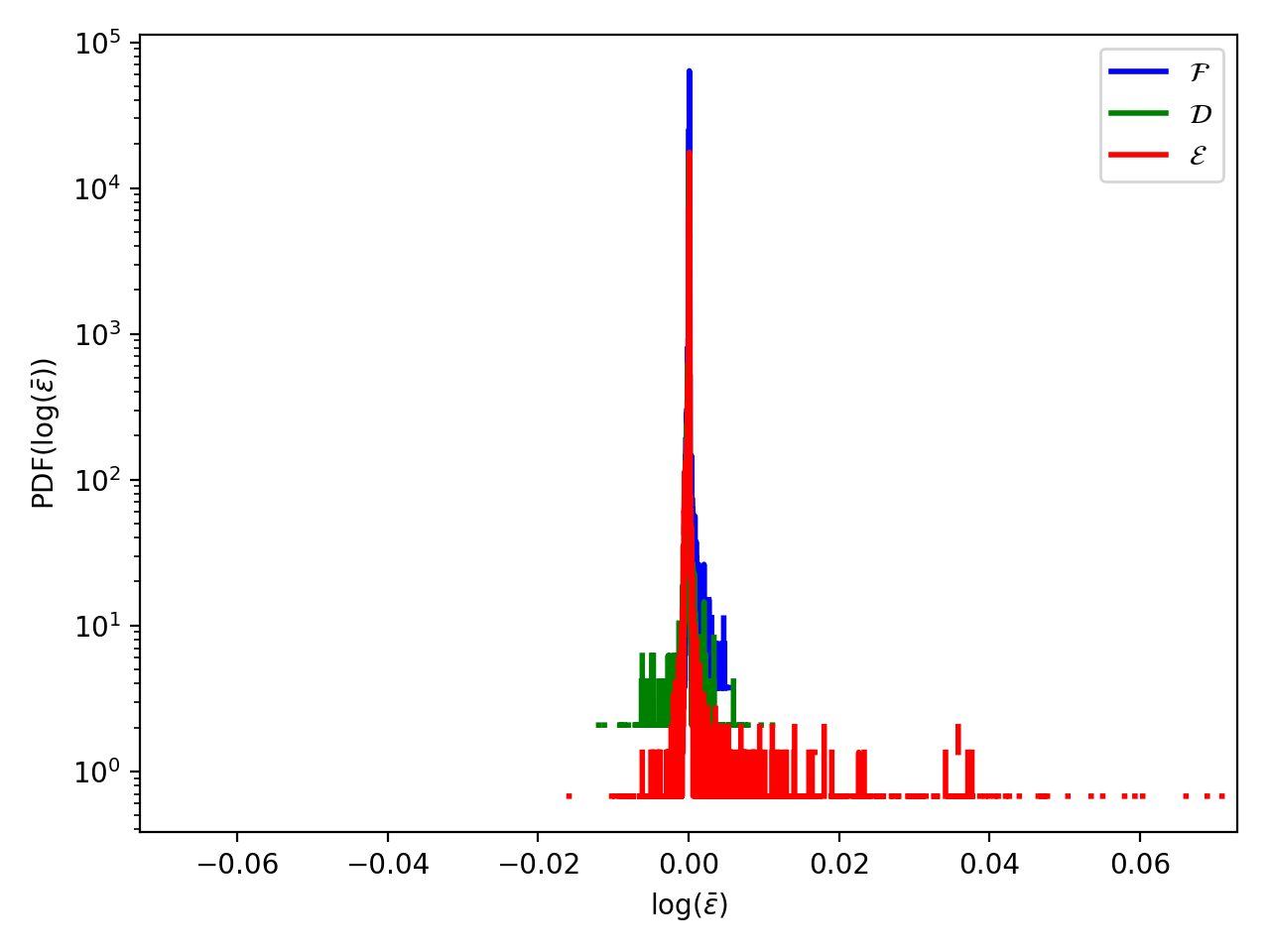}\caption{$\textsc{MRN}$ test  at $t_{\mathrm{core}}+2$~kyr. Probability density function (PDF) of the total dust ratio enrichment $\log(\bar{\epsilon})$ in the core and the fragments, the disk and the envelope. Dust is a very good tracer for the mass here and the dust distribution is almost uniform in the objects considered.}
      \label{fig:fidMRNhisteps}
      \end{figure}
      \begin{figure}[t]
       \centering
          \includegraphics[width=0.5\textwidth]{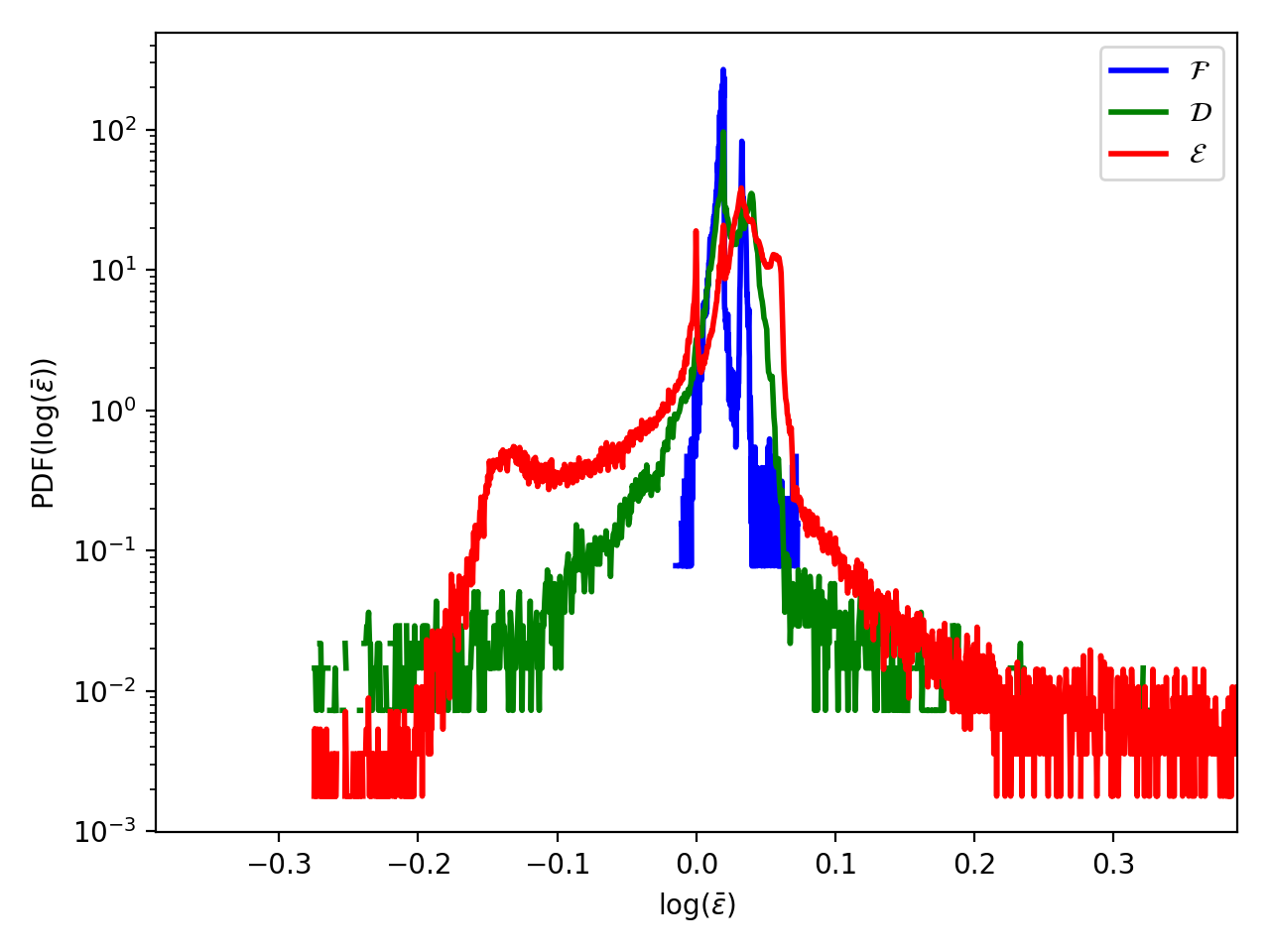}\caption{$\textsc{100micMRN}$ test at $t_{\mathrm{core}}+2$~kyr.  Probability density function (PDF) of the total dust ratio enrichment $\log(\bar{\epsilon})$ in the core and the fragments, the disk and the envelope. Dust is a relatively good tracer for the gas although significant variations of the dust-to-gas are observed.}
      \label{fig:fidmuhisteps}
      \end{figure} 
As seen in Sect. \ref{sec:fid} for the $\textsc{mmMRN}$ model, the differential dynamics between gas and dust during the protostellar collapse depends critically on the grain sizes. Therefore, we perform two simulations $\textsc{100micMRN}$ and $\textsc{MRN}$ with the same set of parameters as in $\textsc{mmMRN}$, but where we vary the maximum grain size. We choose $S_{\mathrm{max}}=100~\micro\meter$ for the $\textsc{100micMRN}$ model (which yields $s_{10}\sim 22.6~\micro\meter$) and $S_{\mathrm{max}}=250~\nano\meter$ for the $\textsc{MRN}$ model (which yields $s_{10}\sim 139~\nano\meter$).

Let us first consider $\textsc{MRN}$,  which is the model that has the smallest $S_{\rm{max}}$. Figure \ref{fig:MRNdens} shows the density of the gas (left) of the least coupled dust species (right)  at $t_{\mathrm{core}}+2$~kyr. Because the coupling between gas and dust is almost perfect, the two maps are indistinguishable by eye. This is expected because the maximum grain size is $\approx 10^{-5}~\centi\meter$ which corresponds to an  initial Stokes number $\mathrm{St}_{0,10} \sim 1.06 \times 10^{-5}\ll 1$(see Sect \ref{sec:ana} for a theoretical justification). As a result, dust is a excellent tracer of the gas in this model. This is illustrated by Fig. \ref{fig:fidMRNhisteps} that shows the probability density function of the dust ratio enrichment $\log(\bar{\epsilon})$ in the core and fragments, the disk and the envelope  at $t_{\mathrm{core}}+2$~kyr. Contrary to the $\textsc{mmMRN}$ model, these PDFs are strongly peaked. The average dust-to-gas ratio integrated over the total dust distribution is $\approx 1\%$ in the core and the fragments, the disk and the envelope. The standard deviation for the dust ratio enrichment ranges between $2 \times 10^{-4}\%$ (in the core) and $1.4 \times 10^{-2}\%$ (in the envelope). For this model, the variations of the dust ratio are very small. Therefore, in absence of coagulation, one may expect that a standard MRN distribution appears to remain extremely well preserved during the protostellar collapse at all scales. 

To investigate an intermediate scenario, we now focus on the $\textsc{100micMRN}$ model. We do not show the density maps in this case due to their strong resemblance with the $\textsc{MRN}$ case.  The PDFs of the dust ratio enrichment $\log(\bar{\epsilon})$ in the core and fragments, the disk and the envelope at $t_{\mathrm{core}}+2$~kyr are shown in Fig. \ref{fig:fidmuhisteps}. In this case, the variations of dust ratio are more significant than in $\textsc{MRN}$.  However, compared to the $\textsc{mmMRN}$ case, these variations still remain quite small.  The average dust-to-gas ratio is $0.0106$ in the core and the disk and $0.0099$ in the envelope. The typical standard deviation for the dust-ratio enrichment are $7 \times 10^{-3}$ in the core, $2.2\times 10^{-2}$ in the disk and $7.8\times 10^{-2}$ in the envelope.  This confirms that the larger the grains are, the more significant the decoupling with the gas is. We note that, for $\textsc{100micMRN}$, it is reasonable to infer the gas density from the dust.

\subsubsection{Thermal-to-gravitational energy ratio}

 \begin{figure}[t!]
       \centering
          \includegraphics[width=0.5\textwidth]{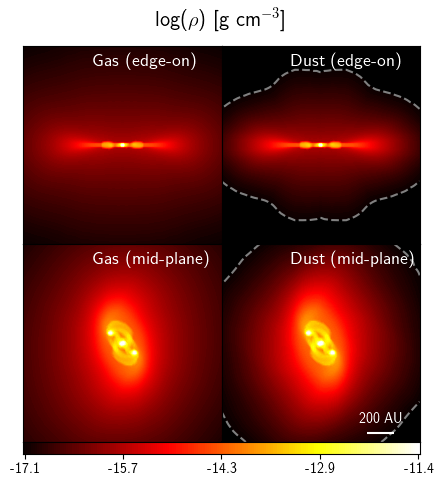}
      \caption{$\textsc{mmMRNa0.25}$ test  at $t_{\mathrm{core}}+2$~kyr ($t_{\mathrm{core}}= 23 $~kyr). Edge-on (top) and mid-plane (bottom) cuts of the gas density (left) and  dust density of the least coupled species (right).  Dust has less time to significantly decouple from the gas than in the $\textsc{mmMRN}$ case. A strong dust depletion is observed in the envelope.}
      \label{fig:MRNdensa025}
      \end{figure}

  \begin{figure}[t!]
       \centering
          \includegraphics[width=0.5\textwidth]{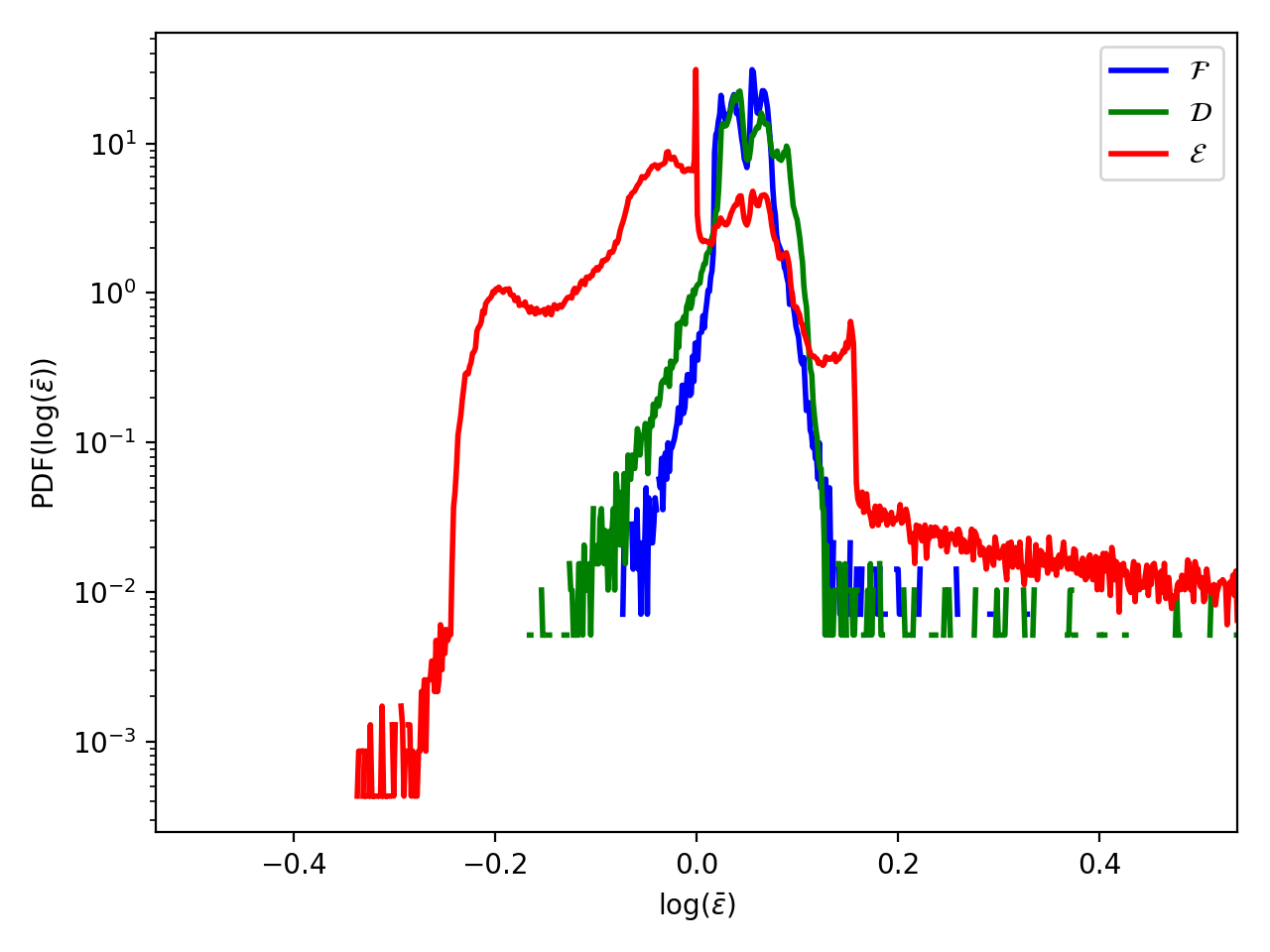}\caption{$\textsc{mmMRNa0.25}$ test  at $t_{\mathrm{core}}+2$~kyr. Probability density function (PDF) of the total dust ratio enrichment $\log(\bar{\epsilon})$ in the core+fragments, the disk and the envelope. Dust is a relatively good tracer for the gas in the dense object although a notable depletion is observed in the envelope.}
      \label{fig:025histeps}
      \end{figure}    
    
The free-fall timescale depends on the ratio between the thermal and the gravitational energy. We therefore present in this section $\textsc{mmMRNa0.25}$, a model similar to the reference case but with  $\alpha= 0.25$.  This parameter is expected to affect strongly the dust dynamics. A lower value of $\alpha$ produces faster protostellar collapses prior to the first core formation due to the smaller thermal support. It thus develops faster high densities regions where dust strongly couples. In addition, a cloud with a smaller $\alpha$ has a smaller initial Stokes number, which means that the dust is also initially better coupled with the gas. The post-core evolution of $\textsc{mmMRNa0.25}$ is different than in  $\textsc{mmMRN}$ in virtue of a smaller initial disk radius.  Smaller disks with a smaller initial value of $\alpha$ are expected as shown by \cite{2016ApJ...830L...8H} (their Eq. 14).

Figure \ref{fig:MRNdensa025} shows the densities of the gas (left) and the least coupled dust species (right)  at $t_{\mathrm{core}}+2$~kyr for $\textsc{mmMRNa0.25}$. Apart from the outer regions that are quite depleted in dust content, we do not see any significant difference between gas and dust. Indeed, the core forms quickly, leaving no time for the differential motion between gas and dust to develop. For a comparison with the fiducial case, we show in Fig. \ref{fig:025histeps} the probability density function of the dust ratio enrichment $\log(\bar{\epsilon})$ in the core and the fragments, the disk and the envelope  at $t_{\mathrm{core}}+2$~kyr. The PDFs are much more peaked in $\textsc{mmMRNa0.25}$ than in  $\textsc{mmMRN}$. The values of the standard deviation of the dust-ratio enrichment are $2   \times 10^{-2} $ in the FHSC and fragments, $ 3 \times 10^{-2}$  in the disk and $0.1 $ in the envelope. This was actually expected as the initial Stokes number scales as $\alpha^{2/3}$ and is thus $\approx 0.6$ times smaller in $\textsc{mmMRNa0.25}$ than in  $\textsc{mmMRN}$.

\subsubsection{Magnetic fields}
\label{sec:bfield}
\begin{figure}[t!]
       \centering
          \includegraphics[width=0.5\textwidth]{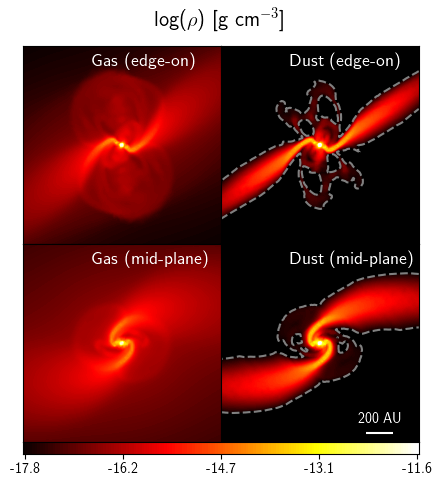}
      \caption{$\textsc{mmMRNmhd}$  at $t_{\mathrm{core}}+2$~kyr ($t_{\mathrm{core}}= 81.1$~kyr). Edge-on (top) and mid-plane (bottom) cuts of the gas density (left) and  dust density of the least coupled species (right).  Dust is significantly decoupled from the gas and concentrate in the high density regions such as the core, the disk, the pseudo-disk and the inner regions of the outflow.}
      \label{fig:idealdens}
      \end{figure}
   
   \begin{figure}[t!]
       \centering
          \includegraphics[width=0.5\textwidth]{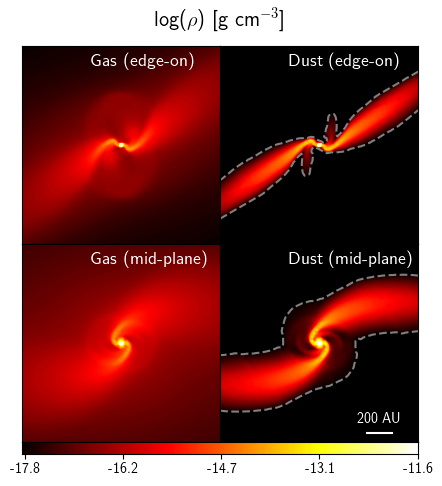}
      \caption{$\textsc{mmMRNnimhd}$  at $t_{\mathrm{core}}+2$~kyr  ($t_{\mathrm{core}}= 81.1$~kyr). Edge-on (top) and mid-plane (bottom) cuts of the gas density (left) and  dust density of the least coupled species (right). As in the ideal case, dust is significantly decoupled from the gas and concentrate in the high density regions such as the core, the disk, the pseudo-disk and the inner regions of the outflow .}
      \label{fig:nimhddensa}
      \end{figure}
 
  \begin{figure}[t!]
       \centering
          \includegraphics[width=0.5\textwidth]{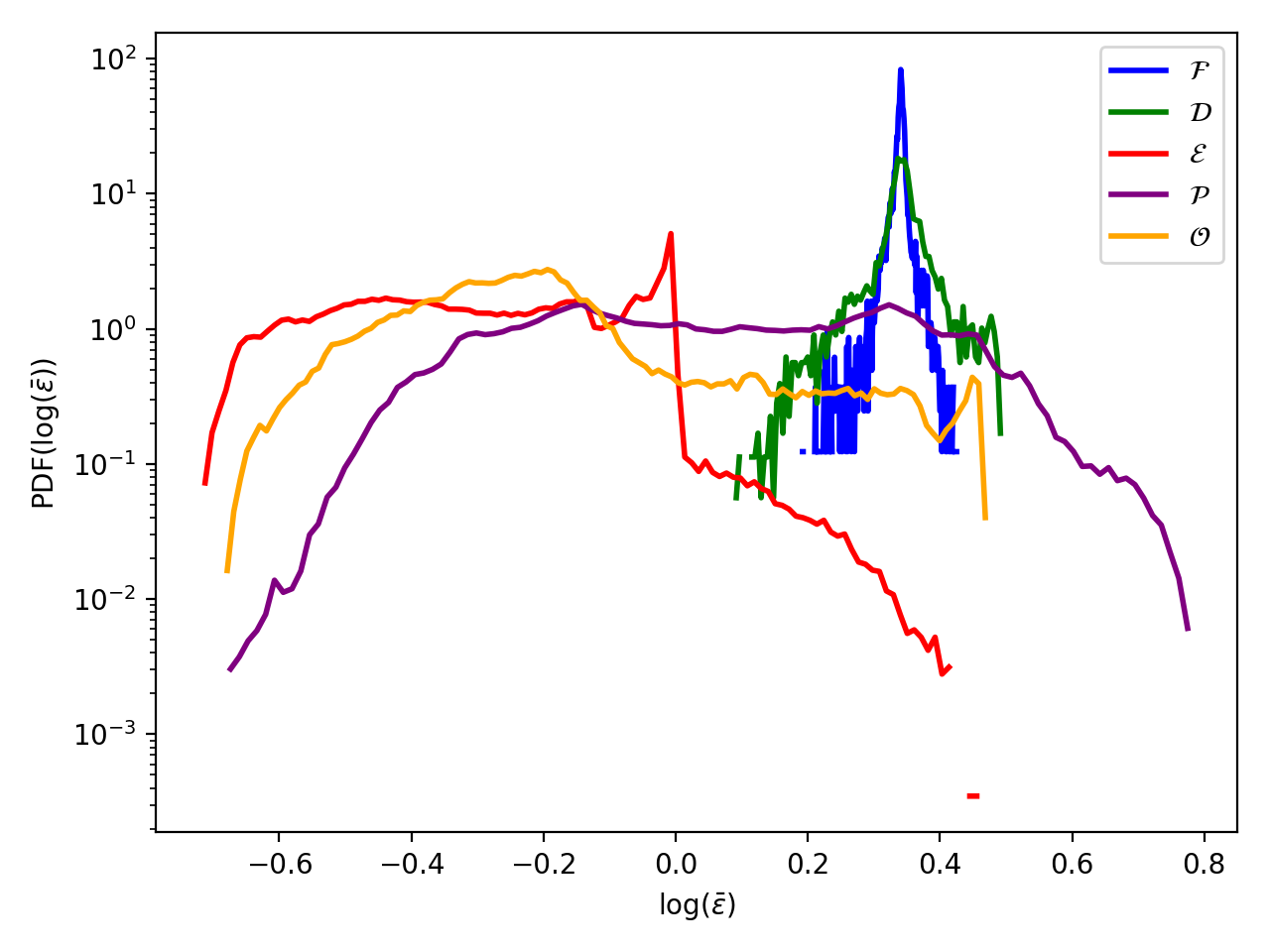}\caption{$\textsc{mmMRNmhd}$ test at $t_{\mathrm{core}}+2$~kyr. Probability density function (PDF) of the dust ratio enrichment $\log(\bar{\epsilon})$  in the core (blue), the disk (green) the pseudo-disk (purple), the outflow (orange)  and the envelope (red).  Dust is not a good tracer for the gas here and the dust distribution is not uniform in the considered objects.}
      \label{fig:idealhisteps}
      \end{figure}   
      
       \begin{figure}[t!]
       \centering
          \includegraphics[width=0.5\textwidth]{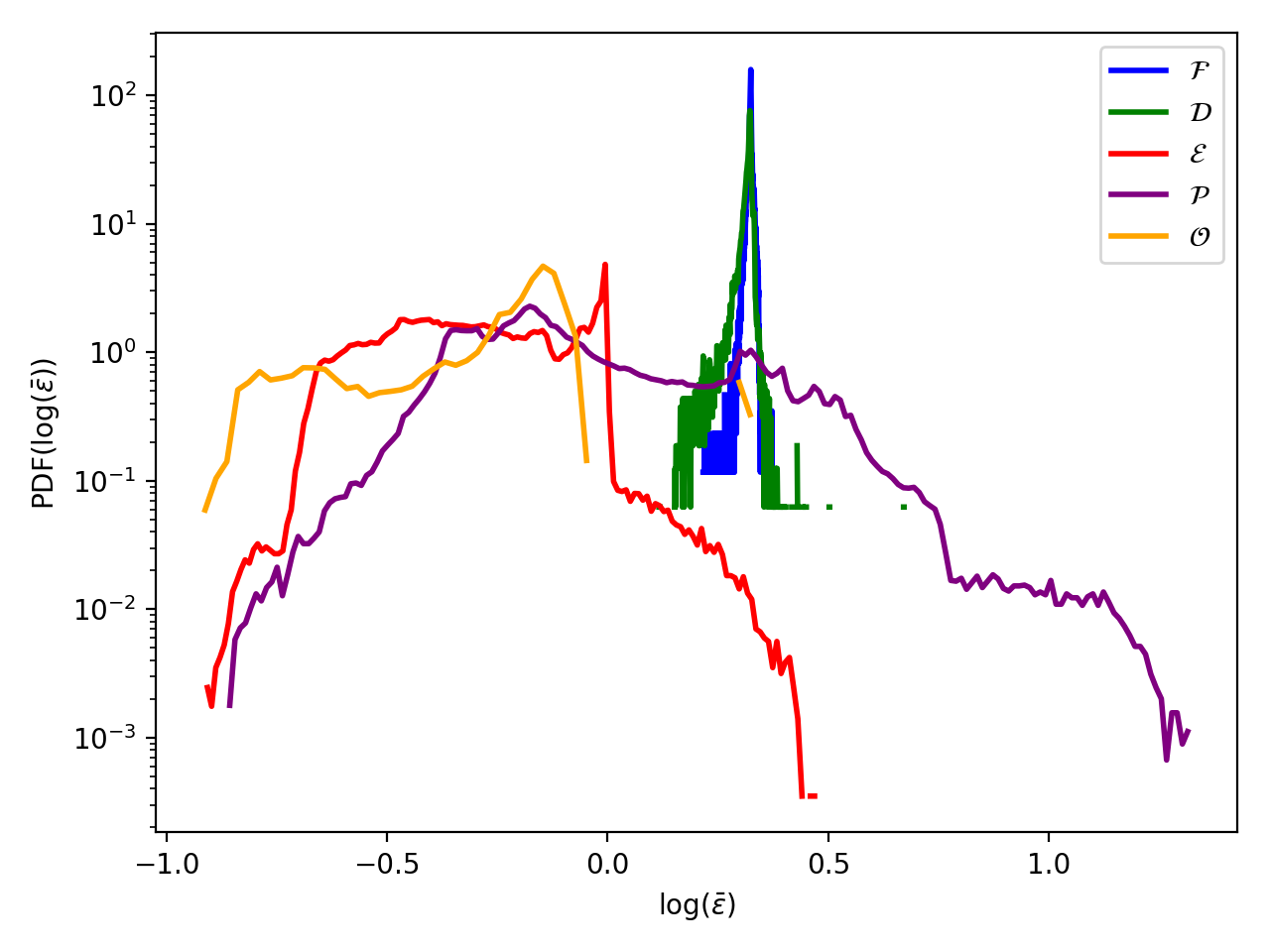}\caption{$\textsc{mmMRNnimhd}$ test   at $t_{\mathrm{core}}+2$~kyr. Probability density function (PDF) of the dust ratio enrichment $\log(\bar{\epsilon})$  in the core (blue), the disk (green) the pseudo-disk (purple), the outflow (orange)  and the envelope (red). Dust is not a good tracer for the gas here and the dust distribution is not uniform in the considered objects.}
      \label{fig:ambihisteps}
      \end{figure}   
      
We now consider the dynamics of neutral grains in collapsing magnetized clouds. The two models are performed with the same parameters as $\textsc{mmMRN}$ but with an initial magnetic field given by $\mu=5$ and a tilt of $40^{\circ}$. For $\textsc{mmMRNmhd}$ we use an ideal MHD solver and for $\textsc{mmMRNnimhd}$ we consider ambipolar diffusion. 
 
Figures \ref{fig:idealdens} and \ref{fig:nimhddensa} show the densities for the gas (left) and the least coupled dust species (right) at $t_{\mathrm{core}}+2$~kyr for the two models $\textsc{mmMRNmhd}$ and $\textsc{mmMRNnimhd}$ respectively.  For both models, the dust is significantly decoupled from the gas and the settling in the core/disk/pseudo-disk is very efficient. As in $\textsc{mmMRN}$, dense regions (disk, core, pseudo-disk and high density regions of the outflow) are prone to be enriched in solid particles while low density regions are depleted (envelope and low density regions of the outflow).  We note that the decoupling is slightly more efficient in these models than in $\textsc{mmMRN}$. This is mostly due to the $10$~kyr difference in free-fall timescale. Although additional decoupling terms due to the magnetic field appear in the dust differential velocity ($ \propto \vec{J}\times \vec{B}$ see Eq. \ref{eq:wkB}) those are negligible compared to the hydrodynamical terms ($\propto \nabla P_{\rm{g}}$) in the decoupling of gas and dust in our magnetised collapse models. 

 We show fig. \ref{fig:idealhisteps} and \ref{fig:ambihisteps} the PDF of the dust ratio enrichment for the different objects  at $t_{\mathrm{core}}+2$~kyr for the  $\textsc{mmMRNmhd}$ and  $\textsc{mmMRNnimhd}$ runs, respectively. Although the shapes of the distributions are different from our fiducial case, we essentially reach to the same conclusion that is a peaked distribution with a significantly large average, indicating a strong initial enrichment, in the [core+disk] system and a broad distribution in the envelope. We note that the average dust-to-gas ratio in the disk and the core are higher in these two models than in the fiducial case. We indeed measure an average dust-to-gas ratio of $\sim 0.022-0.023$ in the disk and the first hydrostatic core for these models. In these magnetic runs, the pinching of the magnetic field lines during the collapse produces a pseudo-disk, which is a dense but not rotationally supported regions. We note that these pseudo-disks have a very broad PDF and show enriched and depleted regions in both the ideal and non-ideal cases. A similar behavior is observed in the magnetically driven outflow for the ideal case, that are dust-rich in dense regions and depleted at low densities. In the  $\textsc{mmMRNnimhd}$, the outflow is less evolved and is mostly dust-depleted similarly to the envelope. 

In summary, neutral dust dynamics in the presence of magnetic fields seems to follow the same general trend as in the hydrodynamical case. Dust collapses faster than the gas and is enriched in the inner regions of the collapse a few thousand years after the first core formation. This enrichment is mainly located in the pseudo-disk (only observed in the magnetized models), the disk, the first hydrostatic core. and the high density regions of the outflow.

\section{Features of dusty collapses}
 \label{sec:FEAT}
\begin{figure}[t!]
       \centering
          \includegraphics[width=0.5\textwidth]{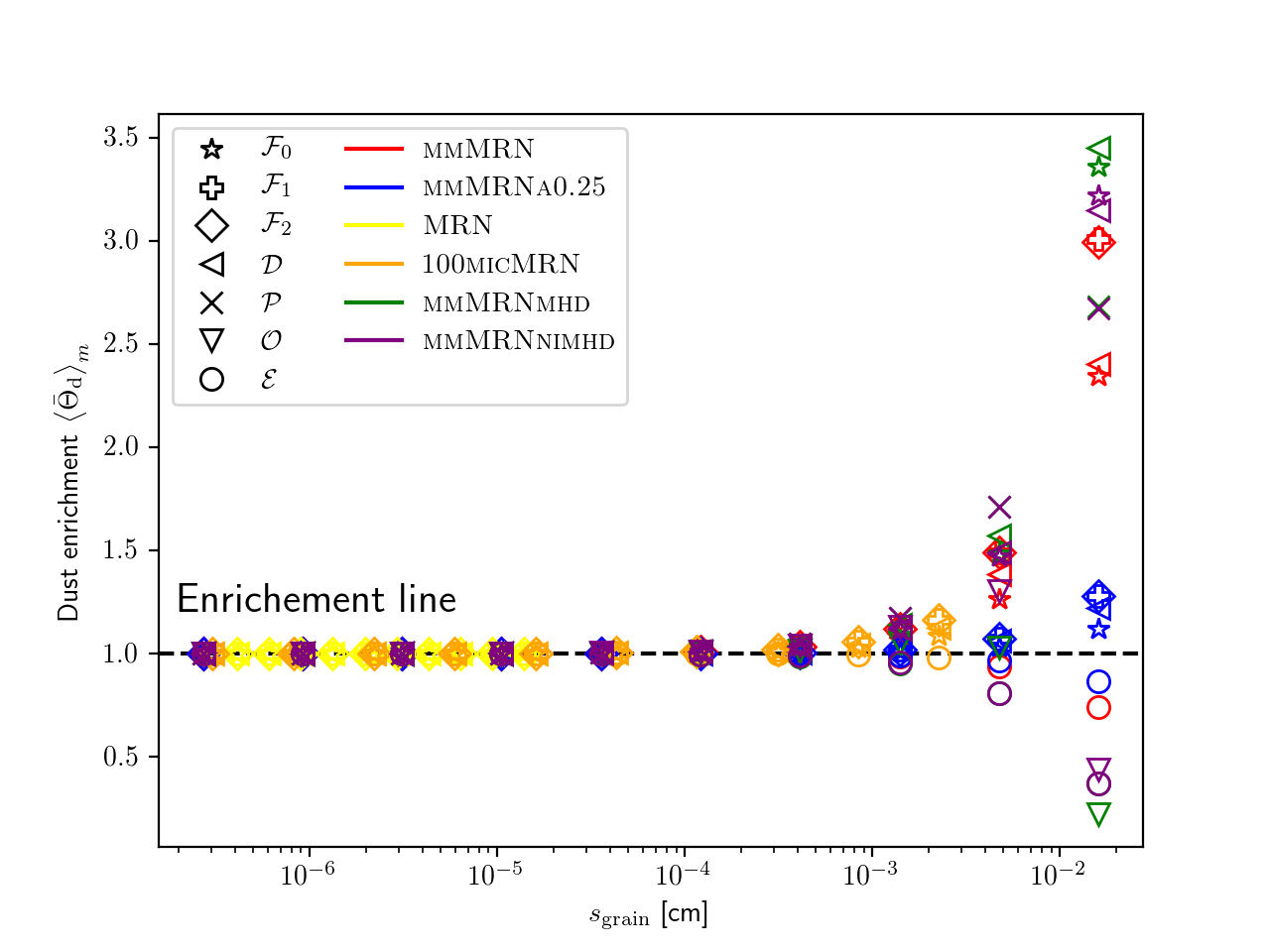}
      \caption{All the models at $t_{\mathrm{core}}+2$~kyr. The dust-to-gas ratio enrichment averaged in mass is shown as a function of the grain size for all the objects. Grain with sizes smaller than $10^{-4}~ \centi\meter$ are almost always perfectly coupled with the gas. For larger sizes, the enrichment is model dependent. Grains with typical sizes larger than $10^{-3}~ \centi\meter$ decouple from the gas.  Dense objects such as the fragments $\mathcal{F}$ or the disk $\mathcal{D}$ and pseudo-disk $\mathcal{P}$ are enriched in dust. Low density objects such as the envelope $\mathcal{E}$ or the outflows $\mathcal{O}$ are depleted in dust. Magnetized models exhibit the stronger decoupling between the gas and the dust.}
      \label{fig:dustenrich}
      \end{figure}

  We summarize here the properties of the dusty collapse in its different regions. Figure \ref{fig:dustenrich} shows the dust-to-gas ratio enrichment averaged in mass as a function of the grain size for all the models and all the objects defined in Sect. \ref{sec:analy}. We refer to the dashed horizontal line as the enrichment line. If an object lies above it, it is enriched in dust during the collapse. If it lies under, it is dust depleted. This information is collected in Table~\ref{tab:corem}.

\subsection{Core and fragments} 

Here we detail the dust and gas properties in the first hydrostatic core and the fragments. In every model, we observe a value of $\left<\Theta_{d,k\leq6}\right>_{m}$ that remains unchanged for all the cores and fragments. Indeed, small grains have short stopping times and remain very well coupled to the gas all along the collapse. Simulations with the largest maximum grain size ($S_{\mathrm{max}}=0.1~\centi\meter$) have the largest dust enrichment. On the contrary, for the $\textsc{MRN}$ model, the dust-to-gas ratio preserves its initial value in all the fragments. Moreover, the dust distribution itself remains extremely well preserved. In the $\textsc{mmMRNa0.25}$ case, the enrichment of the largest dust grains is much less efficient than in $\textsc{mmMRN}$. This is explained by a shorter free-fall timescale and higher initial densities, which implies smaller initial Stokes numbers. In $\textsc{mmMRN}$ $\textsc{mmMRNa0.25}$ and $\textsc{100micMRN}$, the fragments have larger enrichment in large grains ($k>6$, see Tab. \ref{tab:distri} for the corresponding grain sizes) . For example, in the $\textsc{mmMRN}$ case, the dust-to-gas ratio  of the $160~ \micro\meter$ grains is enriched by a factor of $\approx 2.6$ in the central object, and  $\approx 3$  in the fragments. We note that the dust-to-gas ratio enrichment of the first hydrostatic core is stronger in simulations that include magnetic fields than in $\textsc{mmMRN}$. As explained before, magnetic fields bring an additional decoupling in the case of neutral grains which explains in part why the dust enrichment is even stronger in $\textsc{mmMRNmhd}$ and \textsc{mmMRNnimhd} than in $\textsc{mmMRN}$. More importantly, the collapse is longer for these models due to the magnetic support. This leaves more time for dust grains to enrich the central regions. For the fragmenting cases, we note a preferred concentration of dust in the fragments that can be explained by two mechanisms. First, fragments form after the central object and thus stay a longer time in the isothermal phase where dust is less coupled since the temperature is smaller. Second, the dust-rich spiral arms developing through the envelope (see Figs. \ref{fig:dratall10}) are mainly accreted by the fragments (see Fig. \ref{fig:enricht}). This provides an additional channel to enrich the fragments in solids.

\subsection{Disks} 

 We review the disk properties of all the models at $t_{\mathrm{core}}+2$~kyr . Essentially, the values are similar to what is measured in the cores. This is essentially caused by the fact that the dust enrichment happens prior to core formation at low densities (see Fig. \ref{fig:enricht}).  We measure a total dust-to-gas ratio enrichment of $\sim 1.75$ in $\textsc{mmMRN}$, $\sim 1$ in $\textsc{MRN}$, $\sim 1.1 $ in $\textsc{100micMRN}$, $\sim 1.1 $ in $\textsc{mmMRNa0.25}$, $\sim 2.3$ in $\textsc{mmMRNmhd}$ and $\sim 2.1$ in $\textsc{mmMRNnimhd}$. We emphasize once again that the decoupling between the gas and the dust depends strongly on the initial properties of the cloud. We note that in $\textsc{mmMRN}$ the dust ratio is highly non-uniform in the disk (see Figs. \ref{fig:fidhisteps} and \ref{fig:dratall10} for the $\textsc{mmMRN}$ case) or even constant (see Fig. \ref{fig:enricht}). As explained in Sect.\ref{sec:fid}, there is a decrease of $\sim 22\%$ of dust-to-gas ratio in $\textsc{mmMRN}$ between $t_{\rm{core}}$ and $t_{\rm{core}}+4$~kyr. This is most likely due to the fact that the disk is accreting dust depleted material from the envelope. In addition, since dust drifts toward pressure bumps -- or regions where $\vec{J} \times \vec{B} \sim \nabla  P_{\mathrm{g}}$ for magnetic runs -- dust cannot always be used as a direct proxy to trace the gas density. Although, as shown in Fig \ref{fig:mmPress}, the sub-structures seen in the dust originate from the ones in the gas, they may have very distinct morphologies. This is similar to what is found for T-Tauri disks, where gaps could be opened in the dust only \citep{2017MNRAS.469.1932D}.
 Once again, simulations with magnetic fields show a more significant dust enrichment because of the longer timescale of the collapse. We note that the disk masses are however much smaller in the two models where magnetic braking occurs. Indeed, longer integration time is required for the disk to grow significantly \citep{2020A&A...635A..67H}. 
  
\subsection{Pseudo-disks} 

In this section, we describe the principal features of the dusty pseudo-disks that are observed in the two magnetic runs $\textsc{mmMRNmhd}$ and $\textsc{mmMRNnimhd}$. These pseudo-disks are strongly enriched with a total dust-to-gas ratio enrichment of $\sim 2$ for both cases (see values in  Table \ref{tab:corem}). Interestingly, the pseudo-disk is generally more enriched in smaller grains (up to $47~\micro\meter$ grains) than the other objects. This is due to two effects. First, the pseudo-disk is less dense than the central regions of the collapse, namely the disk and the core. This explains why smaller grains are more easily drifting towards it. Second, the strong pressure gradient orthogonal to the pseudo-disk generates a drift from the envelope towards it, even for small grains. Once these grains have reached the pseudo-disk, they couple strongly to the gas while larger grains are able to drift to even deeper regions such as the disk and the core.

\subsection{Outflows}
\label{sec:outfl}
  \begin{figure}[t!]
       \centering
          \includegraphics[width=0.5\textwidth]{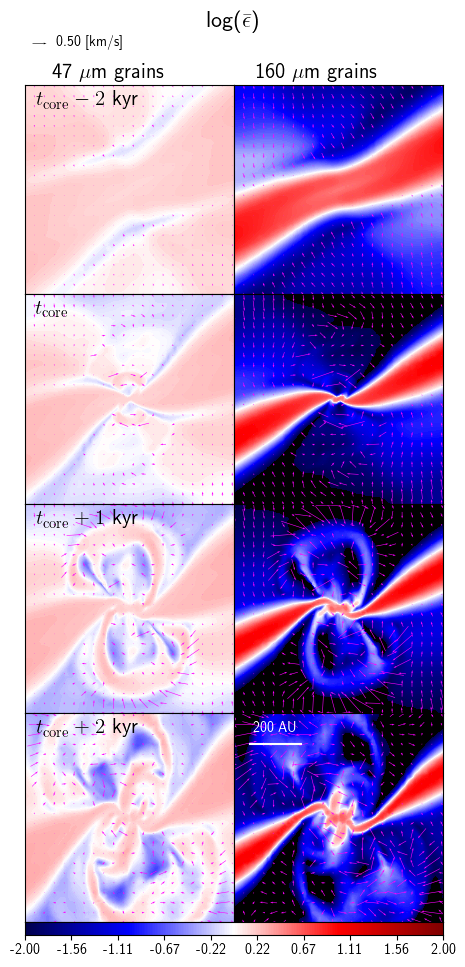}
      \caption{mmMRNmhd. Edge-on view of the relative variations of the dust ratio at four different times for the $47~\micro\meter$ (left) $160~\micro\meter$ grains (right).  The magenta arrows represent the differential velocity with the barycenter. Regions that are dust depleted of more than two orders of magnitude are not displayed (black background).}
      \label{fig:mhdout}
      \end{figure}

We now describe the major features of the dusty outflows that can be observed in the two magnetic runs $\textsc{mmMRNmhd}$ and $\textsc{mmMRNnimhd}$. For $\textsc{mmMRNmhd}$, Fig.~\ref{fig:mhdout} shows the relative variations of the dust ratio at three different times, for the $47~\micro\meter $ (left) and $160~\micro\meter $ (right) grains (9th and 10th bins), respectively. The magenta arrows represent the differential velocity with the barycenter.  These two dust species have a completely different evolution. Indeed, the outflow does not carry a significant quantity of $160~\micro\meter $ grains at  $t_{\mathrm{core}}+ 2~$kyr because they are already strongly depleted in low density regions. Subsequently, the outflow is  strongly depleted in these species with  $\left<\bar{\Theta}_{\rm{d},10}\right>_{m}\sim 0.22$ for $\textsc{mmMRNmhd}$ and $\left<\bar{\Theta}_{\rm{d},10}\right>_{m}\sim 0.44$ for $\textsc{mmMRNnimhd}$.  On the contrary, $47~ \micro\meter $ grains are significantly enriched by a factor 1.04-1.28 at that time. Initially, the outflow is not powerful enough to eject matter from then inner regions and rather collects the grains from the envelope. This explains why the enrichment  measured in the outflow are similar to those measured in the envelope. Interestingly at $t_{\mathrm{core}}+ 2~$kyr, we see that the outflow is well established and starts to carry the inner regions that are denser and more enriched in $160~\micro\meter $ grains. This indicates that outflows provide a channel to re-enrich the envelope  in large grains.

\subsection{Envelope}

 \begin{figure}[t!]
       \centering
          \includegraphics[width=0.35\textwidth]{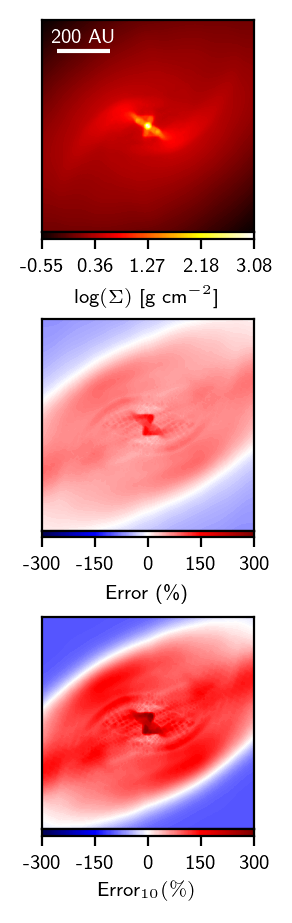}
      \caption{$\textsc{mmMRNmhd}$   at $t_{\mathrm{core}}+2$~kyr. Edge-on view of the total column density $\log(\Sigma)$ (top), the total error  $\mathrm{Err}$  (middle)  and the error when using the largest grains only -- $160~\micro\meter$ in this case (bottom).}
      \label{fig:colmhd}
      \end{figure}

In general, the envelope is dust depleted owing to mass conservation since the inner parts are enriched. For all the models, the small dust grains (with $k\le 6$) do not experience significant dust-to-gas ratio variations. Larger grains can however show significant dust-to-gas ratio variations. We measure a value of $\left<\Theta_{d,10}\right>_{m}$ as low as 0.38 for $\textsc{mmMRNnimhd}$ and $\textsc{mmMRNmhd}$, and 0.73 for  $\textsc{mmMRN}$. This values are averaged over all the envelope, but contrary to the core and fragments, the envelope is very contrasted in terms of density. Among all objects (pseudo-disk excluded), the envelope is experiencing the larger dust-to-gas ratio variations (see for example Fig. \ref{fig:fidhisteps} and \ref{fig:ambihisteps}). This is expected since the density in the envelope is lower than in the other objects. Typically, the depletion in large grain of the envelope increases with a decreasing density and have a larger dust content in their inner regions (see Fig.\ref{fig:mhdout} for both behavior). 

We have shown that dust is not necessarily a good tracer for the gas density, i.e. we find important local variations in the dust distribution in $\textsc{mmMRN}$, $\textsc{mmMRNmhd}$, $\textsc{mmMRNnimhd}$. This has strong consequences for observations, since dust continuum radiation fluxes depend on densities integrated along the line of sight. It is therefore interesting to estimate error that arises when the total column density is estimated from the mass of a single dust bin $k$. We compute this error $\mathrm{Err}_{k}$ from the total column density and the dust column density $\Sigma_{\mathrm{d},k}$ according to 
\begin{eqnarray}
\mathrm{Err}_{k} \equiv \frac{\Sigma -\Sigma_{\mathrm{d},k}/\epsilon_{k,0} }{\Sigma} .
\end{eqnarray} 
The total error $\mathrm{Err}$ is defined the same way but using the total dust column density and initial dust ratio.

Figure  \ref{fig:colmhd} shows an edge-on view of the total column density $\Sigma$ (top), the total error  $\mathrm{Err}$ (middle) and the error estimated by using the largest grains only -- $160~\micro\meter$ in this case (bottom) for the $\textsc{mmMRNnimhd}$ model $2$~kyr after the formation of the first core. The error is large when considering either all the grains (middle) or only the largest ones (bottom). We note that the total column density inferred from the total dust mass would be underestimated in the upper layers and overestimated in the inner regions because dust drifts toward the center of the collapse. The effect is maximal for the largest grains, where the error can reach values as high as $\sim 250 \%$  in the inner envelope. 
  
  \section{Estimate of the dust enrichment}
\label{sec:ana}

Here, we provide a semi-analytic estimate of the dust enrichment occurring during the protostellar collapse. We use it to infer the typical minimal Stokes number above which a given dust enrichment can be reached.

\subsection{Enrichment equation}
In the terminal velocity approximation, and making the assumption that the collapse is is isothermal and purely hydrodynamical, the evolution of the dust ratio for a species $k$ is given by 

\begin{eqnarray}
\label{eq:drat0}
\frac{\mathrm{d}\epsilon_k}{\mathrm{d} t} = -\frac{1}{\rho} \nabla \cdot \left[\epsilon_k \left(\frac{t_{\mathrm{s},k}}{1-\epsilon_k}-\sum_{i=1}^{\mathcal{N}}  \epsilon_i \frac{t_{\mathrm{s},i}}{1-\epsilon_i}\right) c_{\mathrm{s}}^2 \nabla  \rho \left(1-\sum_{j=1}^{\mathcal{N}}  \epsilon_j\right)\right] .
\end{eqnarray}
To provide analytical estimates of dust-ratio enrichment during the collapse, we now neglect the cumulative back-reaction of dust onto the gas. When the cumulative back-reaction is negligible, the evolution of $\bar{\epsilon} = \frac{\epsilon}{\epsilon_0}$ and ${\epsilon}$ are constrained by the same equation, i.e. $\bar{\epsilon}$ does not depend on $\epsilon_{0}$. The dust ratio enrichment is described by the equation

\begin{eqnarray}
\label{eq:drat1}
\frac{\mathrm{d}\bar{\epsilon_k}}{\mathrm{d} t} = -\frac{1}{\rho} \nabla \cdot \left[\bar{\epsilon_k} t_{\mathrm{s}} c_{\mathrm{s}}^2 \nabla  \rho\right] .
\end{eqnarray}
where $\frac{\rm{d}}{\rm{d}t}\equiv \frac{\partial }{\partial t} + \vec{v} \cdot \nabla$. We now use the dimensionless variables
\begin{eqnarray}
\tau =\frac{t}{\tau_{\mathrm{ff},0}} \mathrm{~and~} x =\frac{r}{\lambda_{\mathrm{J},0}}, \nonumber
\end{eqnarray}
where $\tau_{\mathrm{ff},0} = \sqrt{\frac{1}{\mathcal{G}\rho_0}}$ and $\lambda_{\mathrm{J},0}= c_{\mathrm{s}} \tau_{\mathrm{ff},0}$. Equation \ref{eq:drat1} becomes
 \begin{eqnarray}
\label{eq:drat2}
\frac{\mathrm{d}\bar{\epsilon_k}}{\mathrm{d} \tau } = -\mathrm{St}_{k,0}\frac{1}{\bar{\rho}} \nabla_x \cdot \left[\bar{\epsilon_k} \bar{\rho}^{-1}  \nabla_x  \bar{\rho}\right],
\end{eqnarray}
where $\bar{\rho}=\frac{\rho}{\rho_0}$. The term $ \bar{\rho}^{-1} $ appears in the divergence owing to $t_{\mathrm{s},k} \propto \frac{1}{\rho}$.
\subsection{Semi-analytical model}

 Neglecting local variations of $\bar{\epsilon}$ in comparison with local density variations yields
\begin{eqnarray}
\label{eq:dratt}
\frac{\mathrm{d}\bar{\epsilon_k}}{\mathrm{d} \tau } \simeq -\mathrm{St}_{k,0} \bar{\epsilon_k}\frac{1}{\bar{\rho}} \nabla_x \cdot \left[ \bar{\rho}^{-1}  \nabla_x  \bar{\rho}\right].
\end{eqnarray}
We then obtain
\begin{equation}
\label{eq:dratA}
\bar{\epsilon} \left(x, \tau\right) = \chi^{\mathrm{St}_{k,0}},
\end{equation}
where $\chi \equiv e^{-\int_0^{\tau}\nabla_x (  \bar{\rho}^{-1}\nabla_x  \bar{\rho})/\bar{\rho}   \mathrm{d}\tau}$ is independent of the dust properties. Hence, at a given time and position, the dust enrichment varies essentially exponentially with the initial Stokes number. A proper mathematical estimate of the integral quantity is beyond the scope of this study.

\subsection{Estimate in the core}

We can roughly approximate $\chi$ in the core and after a free-fall time $t_{\mathrm{ff},0} = \sqrt{\frac{3 \pi}{32}} \tau_{\mathrm{ff},0}$. An order of magnitude estimate provides 
\begin{equation}
\label{eq:dratchi}
|\mathrm{ln} \left(\chi \right)|\approx \sqrt{\frac{3 \pi}{32}} \frac{ c_{\rm{s}}^2}{\mathcal{G} \rho_{\rm{ad}} r_{\rm{var}}^2} ,
\end{equation}
where $r_{\rm{var}}$ is the typical length at which variations of density become significant. Above $\rho_{\mathrm{ad}}$, the temperature are high and the gas and dust differential dynamics is negligible. A reasonable choice for $\bar{\rho}$ is therefore $\bar{\rho} = \frac{\rho_{\mathrm{ad}}}{\rho_0}$. In the typical condition of a protostellar collapse, one obtains
\begin{equation}
\label{eq:dratchi2}
|\mathrm{ln} \left(\chi \right) |\approx 126  \left(\frac{r_{\rm{var}}}{1 ~\mathrm{AU}} \right)^{-2} \left(\frac{T_{\rm{gas}}}{10~\kelvin}\right).
\end{equation}
We note that taking $r_{\rm{var}}=1$~AU seems reasonable as it is about a tenth of the first-core radius in our models. For $r_{\rm{var}} \approx 5~$AU, we find that $|\mathrm{ln} \left(\chi \right)|\approx 5$. The value of $\chi$ strongly depends on the steepness of the pressure gradients, hence on $r_{\rm{var}}$. This model only provides an rough estimate of $\bar{\epsilon}$ and should not replace either a numerical treatment of the dust or a proper estimate of $\chi$ during the collapse.

\subsection{Comparison with the models}

\begin{figure}[t!]
       \centering
          \includegraphics[width=0.5\textwidth]{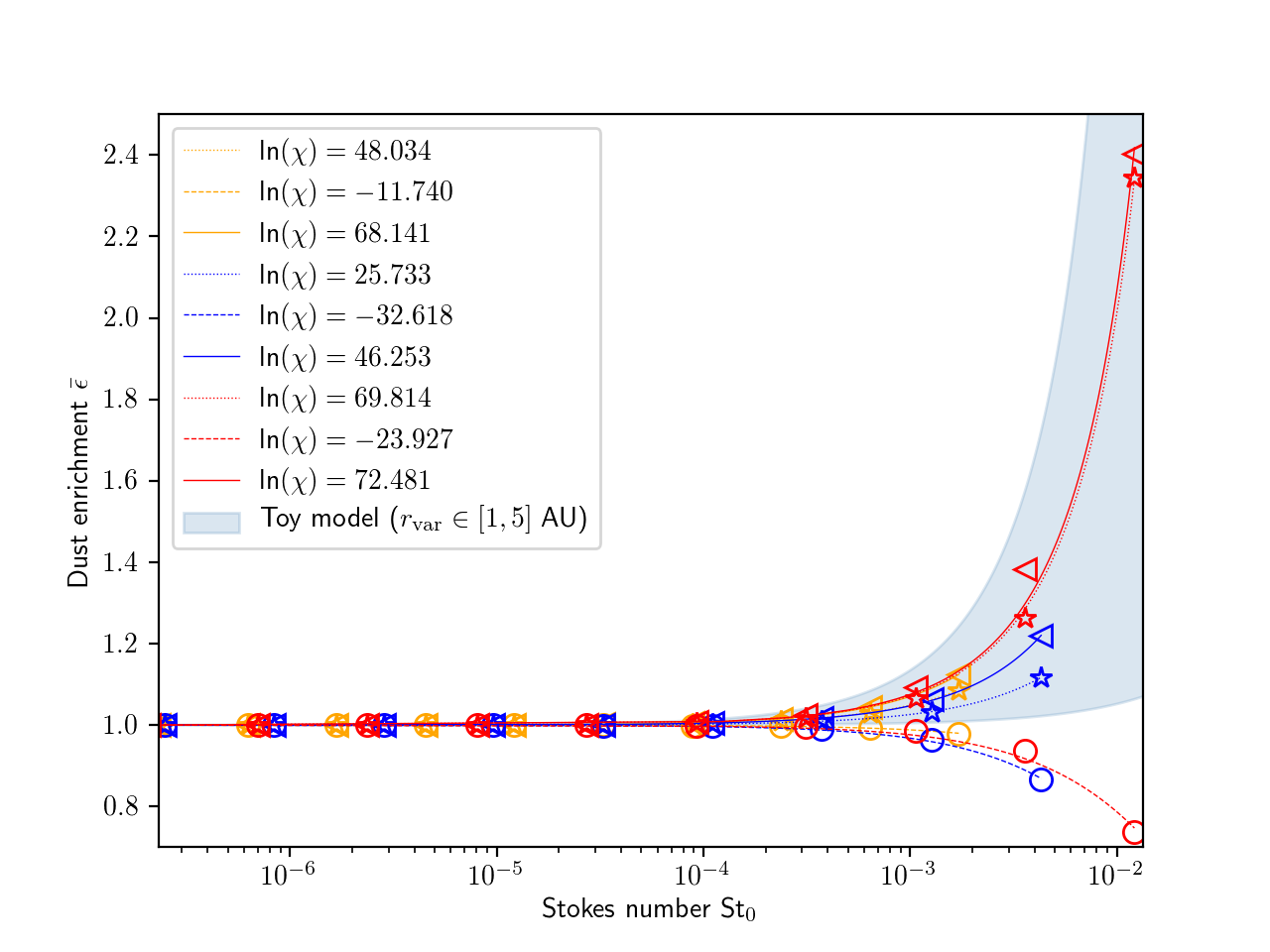}
      \caption{Semi-analytic and measured enrichment in our hydrodynamical models ($\textsc{MRN}$ excluded) against the initial Stokes number. The lines represent the semi-analytical development using a best fit of $\chi$ for the FHSC (dotted), disk (solid) and envelope (dashed), respectively. The extreme values of the enrichment given by our toy model are delimited by the blue areas. The color coding and choice of markers is the same as in Fig. \ref{fig:dustenrich}.}
      \label{fig:enrichthe}
      
      \end{figure}
Figure \ref{fig:enrichthe} shows the dust ratio enrichment as a function of the initial Stokes number for $\textsc{mmMRN}$, $\textsc{mmMRNa0.25}$ and $\textsc{100micMRN}$ for the first core, the disk and the envelope. The dashed, dotted and solid lines represent the values obtained with Eq. \ref{eq:dratA} by fitting the values of $\chi$. Finally, the blue area represents the range of dust enrichment obtained with the two extreme values of $\chi$ estimated in the previous section. We do not show the dust enrichment for $\textsc{MRN}$ as it is clearly negligible (see Fig. \ref{fig:dustenrich}). In addition, we do not display the enrichment in the secondary fragments for the sake of readability.

A fairly good agreement between the fits and the measured dust ratio enrichment is observed in all the regions, especially in the cores. This suggests that the enrichment indeed mostly varies exponentially with the initial Stokes number. We do observe small deviations in the disk and the envelope for $\textsc{mmMRN}$. The non-linear behaviour of Eq. \ref{eq:drat0}, either due to local variations of $\epsilon$ or the cumulative back-reaction of dust on the gas is therefore not completely negligible. In appendix \ref{ap:nback}, we show that the dust-to-gas variations induced by the dust back-reaction are almost negligible and that the discrepancy with the exponential increase of the dust ratio observed is most likely caused by local variation of dust-to-gas ratio, e.g the mixing between depleted material of the envelope and dust rich material from the disk.

Finally, we emphasize that the dust enrichment in all the cores is comprised between the lowest and highest value estimated with our toy-model. This model being quite crude, we acknowledge that it cannot compete with an eventual model based on an accurate estimate of $\chi$.

\subsection{Critical Stokes number}

Assuming that the value of $\chi$ is known, it can be used to determine the critical Stokes number $\mathrm{St}_{\mathrm{crit},\bar{\epsilon}}$ above which some regions can reach a given enrichment $\bar{\epsilon}$. Equation \ref{eq:dratA} can indeed by inverted as
\begin{eqnarray}
\mathrm{St}_{\mathrm{crit},\bar{\epsilon}}= \frac{\mathrm{ln} \bar{\epsilon}}{\mathrm{ln} \chi}.
\end{eqnarray}

Using the value of $\chi$ obtained by fitting our models (Fig. \ref{fig:enrichthe}), we can estimate the typical Stokes numbers needed to get a dust enrichment by a factor of $2$ in the core and the disk is approximately  $\mathrm{St}_{\mathrm{crit},2}\sim 0.01- 0.027$. With our toy model, we find $\mathrm{St}_{\mathrm{crit},2}\sim 0.006- 0.13$. It is a significantly wider range but it contains what is measured with our models. Similarly, we can estimate that to get a dust ratio depletion of $50\%$, the grains must typically have  $\mathrm{St}_{\mathrm{crit},1/2}\sim 0.027-0.11$. In short, it is easier to enrich the disk and the core than it is to deplete the envelope.
\section{Discussion}
\label{sec:DISC}
\subsection{Summary of the models}

 \begin{figure}[t!]
       \centering
          \includegraphics[width=0.5\textwidth]{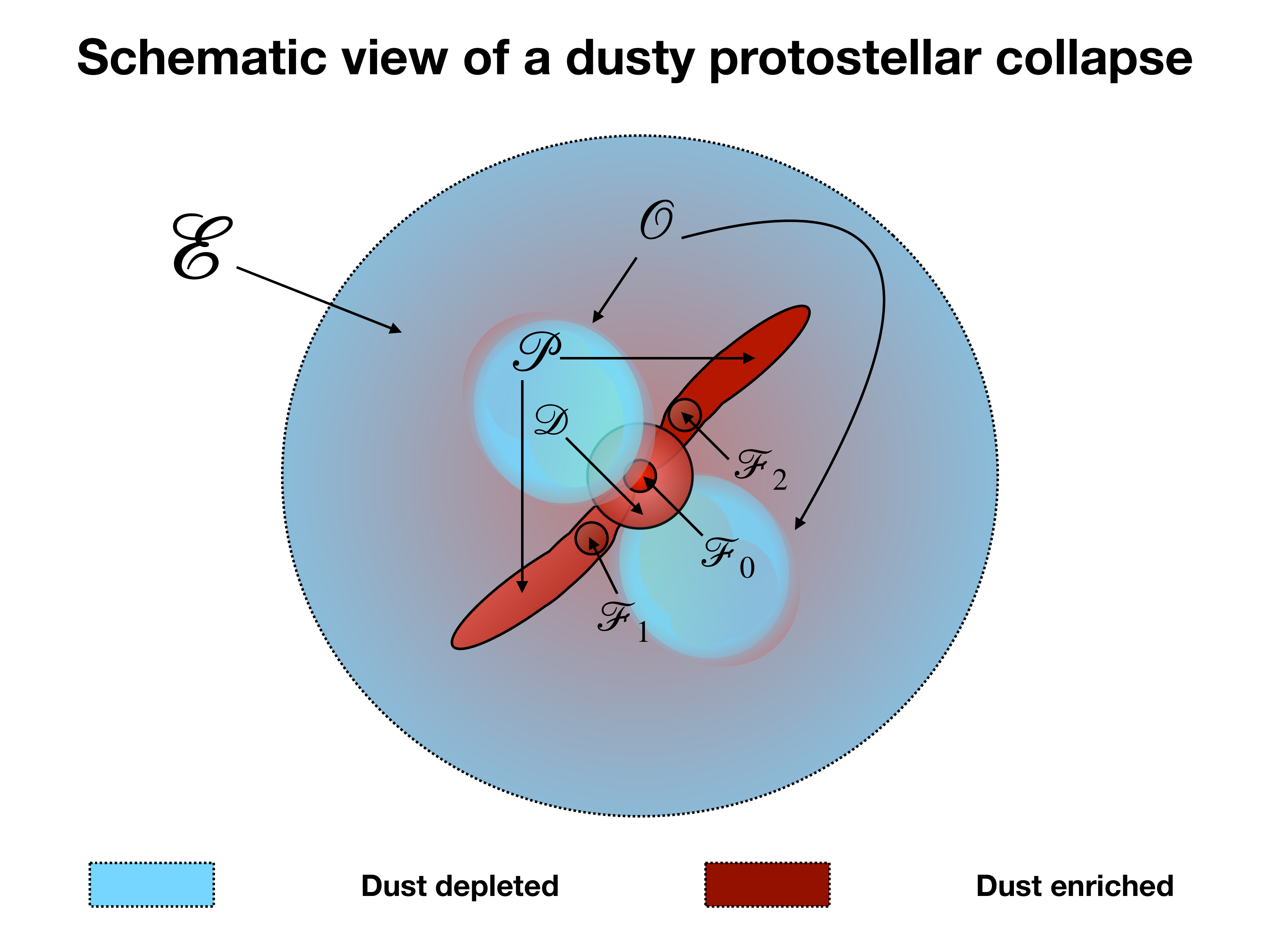}
      \caption{Schematic view of a dusty protostellar collapse. Blue regions are dust depleted and red regions are dust enriched. Typically, the outer regions of the envelope is depleted. The outflow, only observed in magnetic runs, is enriched on its surface and depleted elsewhere. Dense regions, such as the core and  fragments $\mathcal{F}$ (the fragments are observed only in the hydrodynamical case), the pseudo-disk $\mathcal{P}$ (only in magnetic runs) and the disk $\mathcal{D}$ are generally enriched.  The strength of dust decoupling depends on the initial choice of parameters such as the maximum grain size, the thermal-to-gravitational energy ratio or the presence of a magnetic field. This view of a dusty protostellar collapse is simplified and provides only a global sketch of the evolution.}
      \label{fig:dustycoll}
      \end{figure}

We investigated the effect of several parameters on the dust dynamics during the protostellar collapse such as the thermal-to-gravitational energy ratio $\alpha$, the maximum grain size of the dust distribution and the presence of magnetic fields. It appears that the first two parameters are critical for the development of a significant differential dynamics between the gas and the dust during the early phases of the collapse. 

From $\textsc{mmMRN}$, $\textsc{MRN}$ and $\textsc{100micMRN}$, we established that the maximum grain size is a critical parameter for the differential gas and dust dynamics. Typically, if the maximum grain size in the core is smaller than a few microns (see Fig. \ref{fig:dustenrich}), we do not observe significant variations of the dust ratio. On the contrary, when larger grains are considered, the total dust ratio can increase by a factor of $2-3$, even in the very first thousands of years of the protostellar collapse (see $\textsc{mmMRN}$, $\textsc{mmMRNmhd}$ or $\textsc{mmMRNnimhd}$). Hence,  in order to understand the initial dust and gas content of protoplanetary disks, it is crucial to measure accurately the dust size distribution in early prestellar cores. This conclusion is reinforced by recent observations \citep{2019arXiv191004652G} or synthetic observations \citep{2019MNRAS.488.4897V} that seem to probe the existence of $\sim 100~ \micro\meter$ in Class 0 objects. 

For small initial values of $\alpha$, e.g. in $\textsc{mmMRNa0.25}$, the collapse is fast and large grains do not have the time to significantly enrich the core and the disk in one free-fall timescale. Besides, as $\textsc{mmMRNa0.25}$ is set with a higher initial density, dust is initially more coupled with the gas in this particular model. We point out that the efficiency of the dust enrichment relies strongly on the lifetime of low densities regions (see Fig.\ref{fig:enricht}) and on the range of densities experienced during the collapse (see Sect. \ref{sec:ana}). Hence, although $\textsc{100micMRN}$ has a smaller maximum grain size as $\textsc{mmMRNa0.25}$, both models have a similar total dust content by the end of the simulation as the free-fall timescale in $\textsc{100micMRN}$ is longer. The initially properties of the cloud appear to be extremely important to quantify the evolution of the dust distribution during the protostellar collapse. It would be therefore interesting to study the dust collapse of a Bonor-Ebert sphere, since its free-fall timescale is usually longer than the one of the Boss and Bodenheimer test \citep{10.1093/mnras/stt2343}. We leave this proper comparison to further works.

With $\textsc{mmMRNmhd}$ and $\textsc{mmMRNnimhd}$, we investigated the effect of a magnetic field on the dynamics of dust during the protostellar collapse. We qualitatively find similar results as in our fiducial case. Quantitatively, the decoupling between the gas and the dust does however produce more significant variations of the dust-to-gas ratio in the magnetic case. The presence of a dense and stratified pseudo-disk strengthens the envelope and the outflow depletion. This pseudo-disk is consequently strongly enriched in solids. It is in fact almost as enriched as the disk and the first hydrostatic core. However it is much more massive than the disk by the end of the calculation. Therefore, understanding how the pseudo-disk is accreted by the core and the disk is of particular interest and future studies should focus on its long time evolution.
 
For the sake of summarizing, we show in Fig. \ref{fig:dustycoll} a schematic view of a dusty protostellar collapse a few kyr after the formation of the first hydrostatic core. The blue areas depict the dust depleted regions (low density regions of the envelope and outflow) and the red areas represent the regions enriched in dust (cores, disk high density regions of the envelope and outflow, and pseudo-disk). The intensity of the gas and dust decoupling depends naturally of the parameters that we presented earlier in this article. We emphasize that this cartoon illustration is only a simplified picture of a dusty protostellar collapse that does not account for the variability between the models and were we do not quantitatively show the local variations of the dust-to-gas ratio.

\subsection{Comparison with previous works}

Our results are in qualitative agreement with the previous study of \cite{2017MNRAS.465.1089B}, where a decoupling between gas and dust for grains larger than $\approx 100~ \micro\meter$ was also identified. A main difference is that we do not obtain as large dust-to-gas ratio enhancements. Indeed, in \cite{2017MNRAS.465.1089B}, the dust mass is distributed in a single bin of dust with mass of $1\%$ of the mass of the gas. In our \textsc{multigrain} simulations, only a fraction of the dust mass lies in the largest grains, which provides, less significant  dust-to-gas ratio variations as in \cite{2017MNRAS.465.1089B}. This effect was predicted in \cite{2017MNRAS.465.1089B}. We note that $\bar{\epsilon}$ only depends on the initial dust content via the cumulative back-reaction of the dust on the gas. If this back-reaction is neglected as in Eq. \ref{eq:drat1}, the enrichment is independent from the initial value of the dust-to-gas ratio. This allows us a more direct comparison with \cite{2017MNRAS.465.1089B}. In their study, they observe an increase of dust-to-gas ratio of about one order of magnitude for the $100 ~\micro\meter$ grains, which is about 3 times larger than what we observe in $\textsc{mmMRN}$ for example. Using the calibrated value of $\chi$ and Eq.\ref{eq:dratA}, we can estimate that the dust ratio enrichment of $100~\micro\meter$ for grains with $\rho_{\mathrm{grain}} =3~\gram ~\centi\meter^{-3}$ would be $\sim 2.27$ in the first hydrostatic core.  The difference with \cite{2017MNRAS.465.1089B} is likely due to their use of Bonor-Ebert spheres as initial conditions that have a longer free-fall timescale ($\sim~120$~kyr) and because they have lower initial densities ($\sim~10^{-20} \gram~\centi\meter^{-3}$)and therefore larger initial Stokes numbers ($\sim 1$ for $100~\micro\meter$ grains). In 2D simulation of collapsing gravitoviscous protoplanetary disks, \cite{2019A&A...627A.154V,2019A&A...631A...1V,2020A&A...637A...5E} have focused on the evolution of dust including grain growth and fragmentation. Similarly to our models, they observe local variations of the dust-to-gas ratio in the disks. They also found larger dust-to-gas ratios in the inner regions of the collapse and smaller dust-to-gas ratios a few hundreds of AU away from the core. In their high density clumps, they find dust-to-gas ratios between $1.7 \%$ and $2.3\%$ which is quite similar to our findings. Locally, \cite{2019A&A...631A...1V} observe particularly large increase of dust-to-gas ratio in density clumps, which is typically what we observe in our secondary fragments.  

In \cite{2019A&A...626A..96L}, we performed three collapse simulations of non-rotating gas and dust mixture, considering only single dust species ($1~\micro\meter$, $10~ \micro\meter$ and $100~\micro\meter$). In this work, we already observed a significant decoupling occurring for  $100~\micro\meter$ grains. However the increase of dust-to-gas ratio in the core was strong only in the outer regions of the collapse. As previously said, we can estimate that in $\textsc{mmMRN}$ the dust-to-gas ratio enrichment of $100~\micro\meter$ with $\rho_{\rm{grain}}=3~\gram ~\centi\meter^{-3}$ would be about 2.27. It was only $\sim 1.2$ in our first non-rotating spherical collapse calculation, although both models have the same initial $\alpha$.  We interpret this result as an effect of rotation. Firstly, because it slows down the collapse (by a factor $\sim 0.87$ here), which leaves more time for the central regions to be enriched in dust. Secondly, because it generates steeper vertical pressure gradients which allows a more efficient settling of the dust grains. We do not aim to investigate the effect of the initial angular velocity in details since this was done by \cite{2017MNRAS.465.1089B}. The initial angular velocity was found to simply enhance the differential gas and dust dynamics similarly to what the thermal-to-gravitational energy ratio would do. We choose not to explore the impact of grain density because the dependence of the Stokes number in this quantity is the same as for the grain size.

\subsection{Possible implications for planet formation}  
 
The simulations presented in this study consolidate the idea that protostellar collapses may form protoplanetary disks containing $\epsilon_{0} \gtrsim 2-3 \%$ of their mass under the form of solids. 

When the dust ratio $\epsilon_{0}$ is larger than the square of the aspect ratio of the disk $\left(\frac{H}{r}\right)^{2}$ -- even by a tiny amount, grain growth is expected to occur so efficiently that pebbles can decouple from the gas before drifting and falling onto the central star \citep{Laibe2014b}. This condition is likely to be fulfilled as protoplanetary disks typically have $\left(\frac{H}{r}\right)^{2}  \simeq 0.01$ \citep{2010ApJ...723.1241A}. The former condition is strengthened by the fact that back-reaction may also inhibit radial-drift and vertical settling (e.g. \citealt{Kanagawa2017a,Dipierro2018,Lin2019}), and holds until grains fragment. Two scenarios have been debated when fragmentation occurs. In the first scenario, grains may fall onto the central star if the disk does not contain a pressure trap (e.g. \citealt{Brauer2008,Birnstiel2009}). In the second one, dust may exert strong drag onto the gas and powers the development of self-induced dust traps \citep{GLM2017}. The formation of these traps occurs when back-reaction dominates locally over gas viscosity, and may be extremely effective for $\epsilon_{0} \gtrsim 2-3 \%$. In any case, large dust contents favour the formation of planetesimals through the development of the streaming instability (e.g. \citealt{Johansen2007,Johansen2009,Dra2016}). This instability may be even more effective when it develops through unstable epicyclic modes \citep{Jaupart2020}, although peculiar dust distributions may quench it \citep{Krapp2019}.  As argued by \cite{2019A&A...631A...1V}, dust-rich density clumps, similar to the secondary cores in our fragmenting models, could be a favored locus of giant planet formation. They indeed noted both the piling-up of large grains and important growth in these clumps.

In short, a larger initial dust content always favour planet formation in disks, and this may be in a dramatic manner. A quantitative knowledge of the differential dynamics of gas and dust during the protostellar collapse and the initial dust size distribution in prestellar cores therefore appears essential to understand the early stages of planet formation. In that perspective, an extensive study of dust dynamics and coagulation/fragmentation should be done from the scales of molecular clouds and, through the protostellar collapse, up to protoplanetary disks.

   \subsection{Neutral grains approximation}
 \begin{figure}[t!]
       \centering
          \includegraphics[width=0.5\textwidth]{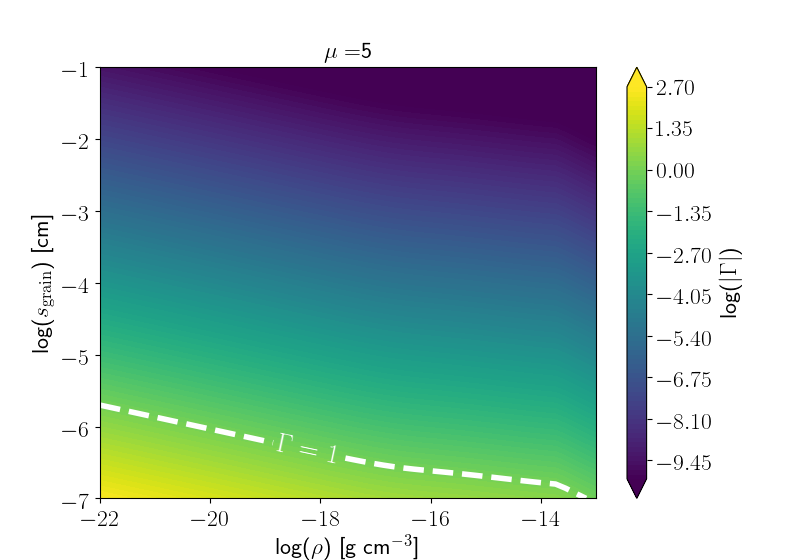}
      \caption{Hall factor as a function of the density and the grain size for $\mu=5$ for  negatively charged grains.  The dashed line denotes the equality between the gyration and the stopping time. }
      \label{fig:StopLor}
      \end{figure}
Grains are likely to be significant or even the main charge carriers during the protostellar collapse \citep{1990MNRAS.243..103U,2016A&A...592A..18M}. Charged dust fluids feel the Lorentz drag $\vec{f}_{\mathrm{L},k}$ in addition to the Epstein drag. The expression of this force is
\begin{eqnarray}
 \vec{f}_{\mathrm{L},k} \equiv \frac{3 Z_k e}{4 \pi \rho_{\mathrm{grain},k} s_{\mathrm{grain},k}^3} \left(\vec{E} + \vec{v}_k \times \vec{B} \right),
\end{eqnarray}
where $Z_k$ is the number of charges on the dust grain $k$, $e =1.6 \times 10^{-20}$~Abc the electron charge and $\vec{E}$ is the electric field. The gyration time of a grain is expressed as
\begin{eqnarray}
  t_{\mathrm{gyr},k}= \frac{4 \pi \rho_{\mathrm{grain},k} s_{\mathrm{grain},k}^3}{3 Z_k e |\vec{B}|} .
\end{eqnarray}

A grain subject to a significant Lorentz drag could preferentially couple to the magnetic field rather than the gas.  In addition, by controlling the Ohmic, ambipolar and Hall resistivities \citep{2009ApJ...693.1895K}, charged grains most likely affect the magnetic and electric fields evolution. To compare the magnetic and neutral drags strength, it is interesting to compare the stopping-to-gyration time ratio also known as Hall factor $\Gamma_k$. In the Epstein regime, this ratio writes 
\begin{eqnarray}
  \Gamma_k \equiv \sqrt{\frac{9 \gamma}{128 \pi}}  \frac{ Z_k e |\vec{B}|}{ \rho c_{\mathrm{s}} s_{\mathrm{grain},k}^2 }.
\end{eqnarray}

To model the Hall factor in a collapsing core, we consider a gas with a barotropic equation of state with $T_{\mathrm{g}}=10$~K, an adiabatic index $\gamma = 5/3$ and negatively charged grains. The grain charge computation follows  \cite{2016MNRAS.457.1037W} and is detailed in Appendix \ref{ap:apA}. We use simple assumptions for the magnetic field, stating that
\begin{eqnarray}
  |\vec{B}|  = \mathrm{min} \left(B_0 {\left(\frac{\rho}{\rho_0}\right)}^{2/3},0.1 \mathrm{G}\right),
  \end{eqnarray}
where $B_0$ is the initial magnetic field, given by $\mu$ and $\rho_0$ the initial core density. The magnetic threshold at $0.1$~G is imposed to reproduce the plateau systematically observed when considering ambipolar diffusion \cite{2016A&A...587A..32M,2016ApJ...830L...8H,2018arXiv180108193V}. One can then show that 
\begin{eqnarray}
  \Gamma_k = \sqrt{\frac{9 \gamma}{128 \pi}}  \frac{ Z_k e}{ \rho c_{\mathrm{s}} s_{\mathrm{grain},k}^2 }\mathrm{min}\left(\frac{{M_{0}}^{1/3}}{\mu \left(\frac{M_{0}}{\Phi}\right)_{\mathrm{c}}} \left(\frac{4 \pi \rho}{3}\right)^{2/3},0.1  \mathrm{G}\right) .
\end{eqnarray}

Fig. \ref{fig:StopLor} shows the absolute value of the Hall factor as a function of the density and the grain size.  For a wide range of grain sizes, the Hall factor is always much smaller than unity. These grains can therefore be considered neutral at least dynamically. We note that $\Gamma_k$ is larger than unity for very small grain ($s_{\mathrm{grain},k} \lessapprox 10^{-6} \centi\meter$). These grains are very well coupled the the gas in the neutral case but, when charged, could experience a strong decoupling with the neutrals. This would occur if the latter are decoupled from the magnetic field e.g., in the non-ideal regime. The Lorentz drag might play a crucial role for the dynamics of very small grains in star formation but is probably not very important for the large grains that we observe to decouple in our models. Efforts to study the dynamics of such grains have been made in the past \citep{2007A&A...476..263G,2018MNRAS.479.4681H} and should be extended to dusty collapses in future studies. We emphasize that the electromotive term in the Lorentz force applies only on the charged species and not the barycenter and might play a very important role in the decoupling between the charged grains and the neutrals (barycenter) even when $\Gamma_k <1$.

 \subsection{Caveat: Coagulation/fragmentation during the collapse}
Dust coagulation has been neglected during this study. It may however affect strongly dust evolution during the collapse since dust decoupling depend on grain sizes. Following \cite{1985prpl.conf..621D}, one can estimate the coagulation timescale $t_{\mathrm{coag},i,j}$ between two dust phases $i$ and $j$ due to their relative motion within the collapsing clouds as 
 \begin{eqnarray}
  t_{\mathrm{coag} ,i,j} = \left[\frac{3}{4} \rho_i \frac{(s_{\mathrm{grain},i}+s_{\mathrm{grain},j})^2}{\rho_{\mathrm{grain},i}s_{\mathrm{grain},i}^3} |\vec{v}_i-\vec{v}_j| \right]^{-1} .
  \label{eq:tcoag}
 \end{eqnarray}
Assuming the same density for all the grains, neglecting cumulative back-reaction effects and magnetic fields, and considering an isothermal collapse, differential velocities can be estimated from the diffusion approximation as
 \begin{eqnarray}
   |\vec{v}_i-\vec{v}_j| \sim \rho_{\mathrm{grain}}| s_{\mathrm{grain},i}-s_{\mathrm{grain},j}| c_{\mathrm{s}} \frac{|\nabla \rho|}{\rho^2}.
 \end{eqnarray}
 
 As said in Sect. \ref{sec:ana}, the density profile of the free-falling material can be approximated as a power law with an exponent $\zeta=-2$ \citep{1969MNRAS.145..271L}. In this case, one obtains at a distance $r$ from the central region
 \begin{eqnarray}
  t_{\mathrm{coag} ,i,j} = \left[\frac{3 |\zeta| }{4 r} \epsilon_i (1+q_{i,j}) (1-q_{i,j}^2)  c_{\mathrm{s}} \right]^{-1}.
  \end{eqnarray}
  where $q_{i,j}\equiv\frac{s_{\rm{grain},j}}{s_{\rm{grain},i}}$ is the ratio between the grain sizes. In the limit $q_{i,j} \ll 1$, we find 
  \begin{eqnarray}
  t_{\mathrm{coag} ,i,j} \sim 170~\mathrm{kyr} \left(\frac{\epsilon_j}{10^{-2}}\right)^{-1}\left(\frac{r}{100~\mathrm{AU}} \right)\left(\frac{c_{\mathrm{s}}}{0.19 ~\kilo\meter~\second^{-1}} \right)^{-1}.
  \end{eqnarray}
 Growth induced by the relative dynamics between dust grains is therefore expected to be not very efficient during protostellar collapse away from the core but could be non-negligible in the inner regions. At $r=10$~AU,   $t_{\mathrm{coag} ,i,j} \sim 17~\mathrm{kyr}$, which is quite smaller than the typical free-fall timescale of a prestellar core. 
 
 We note that growth could also be enhanced by the turbulence, the Brownian motions of dust grains or focalization due to grain charges \citep{2008ARA&A..46...21B}.  In the case of Brownian motions, for example, assuming that the grain temperature is equal to the gas temperature, the differential velocity for two grains of different mass can be expressed as \citep{2016SSRv..205...41B}
  \begin{eqnarray}
  |\vec{v}_i-\vec{v}_j| = \sqrt{\frac{8 k_{\rm{B}} T_{\rm{g}}}{\pi}} \sqrt{\frac{m_{\mathrm{grain},i}+m_{\mathrm{grain},j}}{m_{\mathrm{grain},i}m_{\mathrm{grain},j}}},
  \end{eqnarray}
  hence, assuming again that  $q_{i,j} \ll 1$ the coagulation timescale writes as 
   \begin{eqnarray}
  t_{\mathrm{coag} ,i,j} = \left[\frac{3}{4} \rho_i \frac{1}{\rho_{\mathrm{grain},i}s_{\mathrm{grain},i}} \sqrt{\frac{8 k_{\rm{B}} T_{\rm{g}}}{\pi m_{\mathrm{grain},j}}} \right]^{-1},
  \label{eq:tcoag2}
    \end{eqnarray}
 We now consider a region of density $10^{-12}~ \gram~\centi\meter^{-3}$ and  temperature of $10~\kelvin$. Assuming $s_{\mathrm{grain},j}=0.1~\micro \meter$ and $s_{\mathrm{grain},i}=100~\micro \meter$, we get $t_{\mathrm{coag} ,i,j}\sim 240 ~\mathrm{kyr}$. We note that, in the case of the Brownian motions, the coagulation timescale depends on the grains size of both species. If we now consider $s_{\mathrm{grain},j}=0.01~\micro \meter$, $t_{\mathrm{coag} ,i,j}\sim 7 ~\mathrm{kyr}$.
 
 In short, we are tempted to say that in the presence of large grains, very small grain could be efficiently removed during the collapse in high density regions. Coagulation should therefore be included in future studies.  We admit however that the presence of such large grains during the early phases of the protostellar collapse is still under debate. It is indeed unclear how these large dust grains can overcome the fragmentation barrier, as the differential velocities between two dust species is typically quite large, e.g up to a few $\approx 0.1 \kilo\meter~\second^{-1} $ in the envelope of $\textsc{mmMRN}$ in the case of the two least coupled species. Typically, the velocity above which fragmentation can occur is thought to be about a few $ 10~\meter~\second^{-1}$ \citep{2008ARA&A..46...21B} although it could be higher \citep{2014ApJ...783L..36Y}. Large grains could overcome the fragmentation barrier because the fragmentation timescale is typically equivalent to  coagulation timescale \citep[e.g. ][]{GLM2017}. Fragmentation could simply not have the time to occur during the collapse, especially considering that large grains quickly drift toward regions of high density where their drift velocity is typically around a few meter per seconds. Let us now estimate the typical fragmentation radius $r_{\rm{frag}}$, which corresponds to the distance from the center at which the fragmentation velocity $v_{\rm{frag}}$ is equal to the dust drift velocity. It can be obtained by solving

 \begin{eqnarray}
    \rho_{\mathrm{grain}}s_{\mathrm{grain}} c_{\mathrm{s}} \frac{|\nabla \rho|}{\rho^2} =v_{\rm{frag}} 
 \end{eqnarray}
 
Let us now assume first core formation with a core profile of the type $\rho = \frac{\rho_{\mathcal{F}}}{1+\left(\frac{r}{r_{\mathcal{F}}}\right)^2}$.We can show that
 \begin{eqnarray}
 r_{\rm{frag}} = \frac{1}{2} \frac{r_{\mathcal{F}}^2}{s_{\mathrm{grain}}} \frac{\rho_{\mathcal{F}}}{\rho_{\mathrm{grain}}} \frac{v_{\rm{frag}}}{c_{\mathrm{s}}}.
 \end{eqnarray}

Assuming $\rho_{\mathcal{F}} =10^{-11} \gram~\centi\meter^{-3}$, $r_{\mathcal{F}} = 10~$AU, $\rho_{\mathrm{grain}} = 1 \gram~\centi\meter^{-3}$, $s_{\mathrm{grain}}=160~\micro \meter$ and $T_{\mathrm{g}} =10~\kelvin$

\begin{eqnarray}
 r_{\rm{frag}} &=&  2.5 \times 10^4~{\mathrm{AU}}   \left(\frac{r_{\mathcal{F}}}{10~{\mathrm{AU}}}\right)^{2} \left(\frac{\rho_{\mathcal{F}}}{10^{-11} \gram~\centi\meter^{-3}}\right) \left(\frac{c_{\mathrm{s}}}{0.19 ~\kilo\meter~\second^{-1}}\right) \nonumber \\ & &\left(\frac{v_{\rm{frag}}}{10 ~\meter~\second^{-1}}\right) \left(\frac{s_{\mathrm{grain}}}{160~\micro \meter}\right)^{-1} \left(\frac{\rho_{\mathrm{grain}}}{1~\gram~\centi\meter^{-3}}\right)^{-1}.
 \end{eqnarray}

 The fragmentation radius is significantly larger than the scales that are involved during the first protostellar collapse. We note a quadratic dependency with respect to the first core radius. For a core of size $1~{\mathrm{AU}}$, one finds $  r_{\rm{frag}} =  2.5 \times 10^2~{\mathrm{AU}}$. Hence, fragmentation may be relevant for small cores and should be investigated in further studies.
\section{Conclusion and perspective}
\label{sec:CONCLU}

In this study, we presented the first \textsc{multigrain} and non-ideal MHD simulations of dusty protostellar collapses using the new dust dynamics solver of  \texttt{RAMSES}. We presented six \textsc{dustycollapse} simulations with a simultaneous treatment of $10$ dust species with different sizes. In these simulations, we investigate the impact of the maximum grain size, the thermal-to-gravitational energy ratio and the presence of magnetic fields on the dynamics of the  \textsc{dustycollapse}. We summarize below our principal findings:
\begin{enumerate}
\item Small grains with sizes less than a few $10 ~\micro\meter$ are strongly coupled to the gas during the protostellar collapse. On the contrary, grains larger than $\sim 100~ \micro\meter$ tend to decouple significantly. 

\item When the first hydrostatic core forms, high density regions -- the core, the fragments, the disk and the pseudo-disk -- are enriched in dust by a typical factor of two, whereas low density regions -- the envelope and the outflow -- are strongly dust-depleted.

\item Dust is not necessarily a proxy for gas during the collapse. Inferring gas densities from dust is found to potentially lead to extremely large errors (up to $\sim 250 \%$).

\item A standard MRN grain size distribution with a maximum grain size of $250 \nano\meter$ is however extremely well preserved during the protostellar collapse in absence of coagulation.

\item Dust dynamics is strongly affected by the initial cloud properties. Variations of the dust-to-gas ratio reach the largest values when the free-fall is long and the initial density is low. An additional decoupling occurs for neutral grains in the presence of  magnetic fields because collapse proceeds over longer timescales.

\item With a semi-analytical model, we show that the dust-ratio varies exponentially with the initial Stokes number during the collapse. More precisely, we have shown that it can be expressed as $\epsilon_0\,\chi^{\mathrm{St_0}}$ where $\chi$ is a dimensionless function of the time independent on the dust properties. Fitting the values of $\chi$ gives a very good agreement between the semi-analytical model and the measured dust-ratios in the range of Stokes number considered in our simulations.

\item Using the calibration of $\chi$ with the results of our model, we show that a Stokes number of at least $0.01$ is required to enrich the core  and the disk in a dust species by a factor of $2$. Similarly we show that grains with  $\rm{St_0} \gtrapprox 0.0027$ can potentially be depleted by a factor of $2$ in the envelope after the first core formation.

\end{enumerate}

Dust evolution during the protostellar collapse could have serious consequence on the initial state of protoplanetary disks and the further formation of the planets. In the future, substantial efforts should be made to include the dynamics of charged dust grains during the protostellar collapse, since the Lorentz drag cannot necessarily be neglected for small grains. Coagulation and fragmentation of dust grains should also be considered to investigate more realistically dust evolution during the star formation process.

\begin{appendix}

\section{Impact of velocity regularization}
\label{ap:apA0}

    \begin{figure}[t!]
       \centering
          \includegraphics[width=0.5\textwidth]{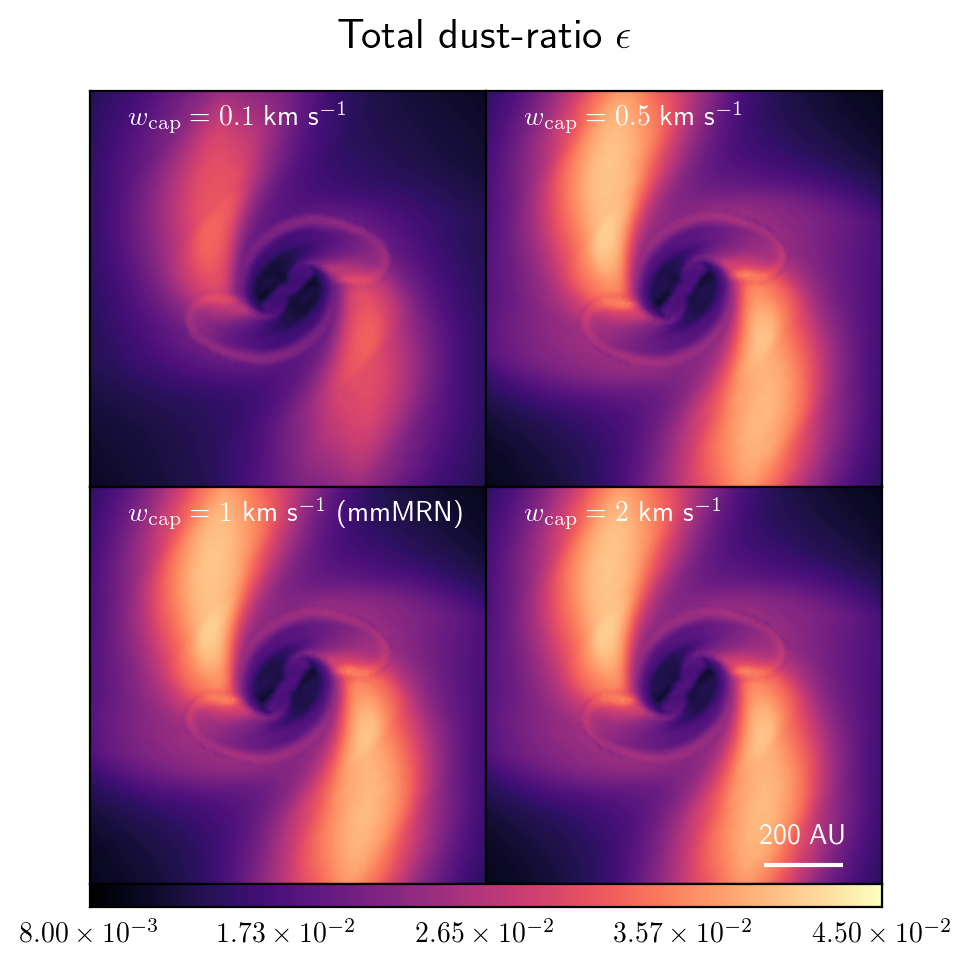}
      \caption{Fiducial model for various maximum dust differential velocity $w_{\mathrm{cap}}$. Mid-plane view of the total dust-ratio at the FHSC formation. (Top-left)  $w_{\mathrm{cap}}=0.1$~ $\kilo \meter~ \second^{-1}$, (Top-right)  $w_{\mathrm{cap}}=0.5$ $\kilo \meter~\second^{-1}$, (Bottom-left)  $w_{\mathrm{cap}}=1$ $\kilo \meter~\second^{-1}$, (Bottom-right)  $w_{\mathrm{cap}}=2$ $\kilo \meter~\second^{-1}$. }
      \label{fig:MRNreg}
      \end{figure}
As explained in Sect.~\ref{sec:regul}, the simulations of our fiducial model have been performed with various values for the maximum gas and dust differential velocity $w_{\mathrm{cap}}$. We investigated the effect of varying $w_{\rm cap}$ by performing complementary simulations with $w_{\mathrm{cap}}=0.1$, $0.5$, $1$ and $2$ $\kilo \meter~\second^{-1}$. We attempted to simulate an additional model with $w_{\mathrm{cap}}=10$ $\kilo \meter~\second^{-1}$ but this led to numerical instabilities due to unrealistically large dust velocities at the accretion shock, where the diffusion approximation is not valid. 

In Fig.~\ref{fig:MRNreg}, we show a face-one view of the dust-ratio at the time of the FHSC formation for these 4 models. With $w_{\mathrm{cap}}=0.5-2~\kilo \meter~\second^{-1}$ we essentially find the same results as in the fiducial case that has $w_{\mathrm{cap}}=1$ $\kilo \meter~\second^{-1}$. Having $w_{\mathrm{cap}}=0.1~\kilo \meter~\second^{-1}$ however appears to be a too extreme choice. It indeed suppresses most of the initial dust enrichment that is due to the decoupling between the gas and the dust in the envelope. Having   $w_{\mathrm{cap}}=1~\kilo \meter~\second^{-1}$ appeared to be a reasonable choice to cope with time stepping and physical constraints.

\section{Non-linear dust enrichment: neglecting back-reaction}

\label{ap:nback}

\begin{figure}[t!]
       \centering
          \includegraphics[width=0.5\textwidth]{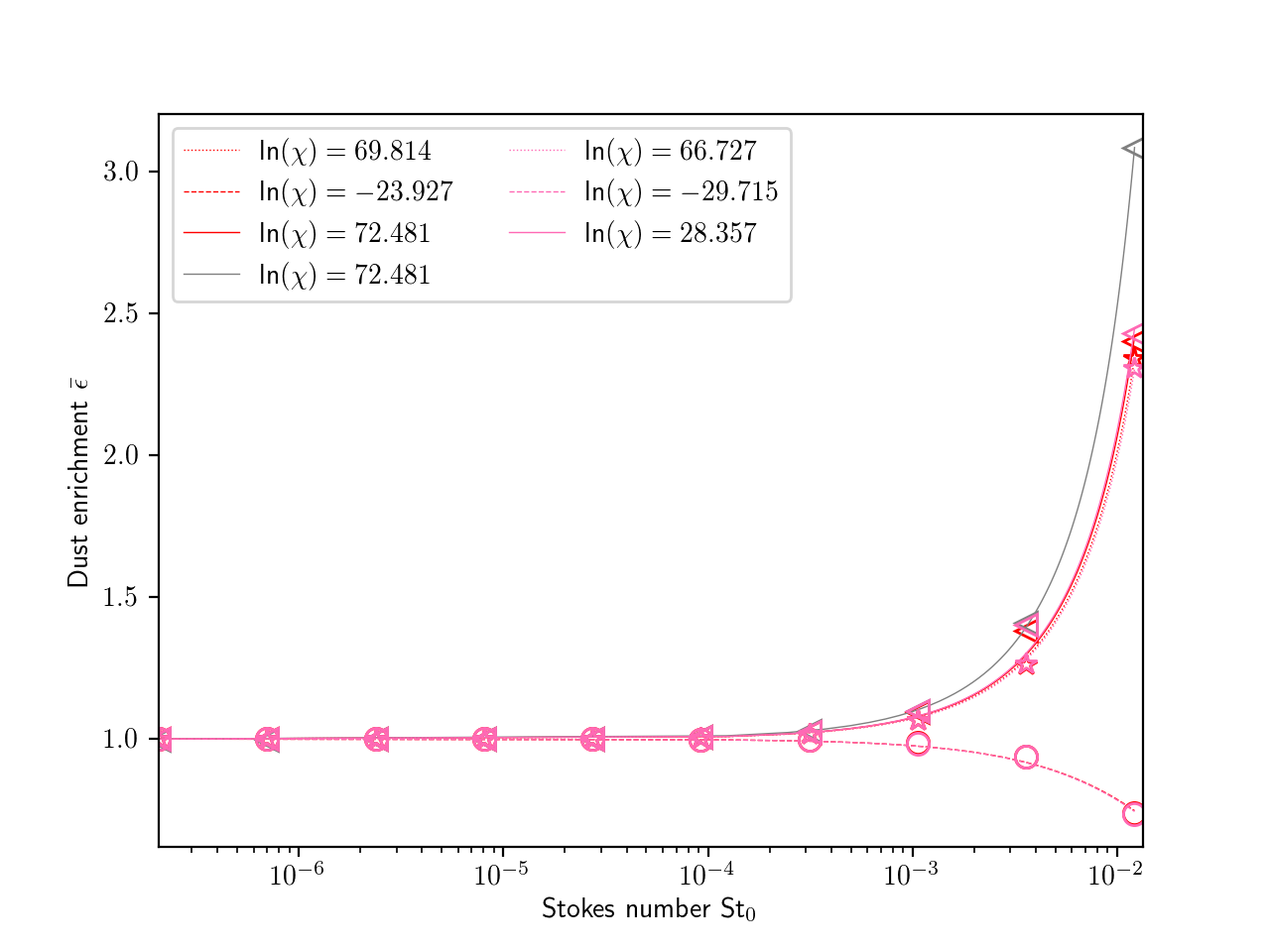}
      \caption{Semi-analytic and measured dust-ratio enrichment in  $\textsc{mmMRN}$ (red) and  $\textsc{mmMRNeps1e-7}$ (pink) in the core (stars), disk (triangles) and envelope (circles) at $t_{\mathrm{core}}+2~$kyr against the initial Stokes number.  The lines represent the semi-analytical development using a best fit of $\chi$ for the FHSC (dotted), disk (solid) and envelope (dashed), respectively. We display the same information for $\textsc{mmMRN}$ at $t_{\mathrm{core}}$  in the disk (grey).}
      \label{fig:enrichback}
      
      \end{figure}
As stated in Sect. \ref{sec:ana}, Eq.\ref{eq:drat2} is valid as long as local dust-ratio variations and the cumulative back-reaction of the dust can be neglected. In this approximation, the dust ratio enrichment increases exponentially with the Stokes number. As pointed out in Sect. \ref{sec:ana}, we observe deviations from the exponential law in the disk and the envelope in $\textsc{mmMRN}$, while it is very well verified in the core. 

To ascertain whether these deviations are due to back-reaction  or not, we perform an additional simulation similar to the fiducial one, but with $\theta_{\mathrm{d},0}=10^{-7}$ ($\textsc{mmMRNeps1e-7}$). For $\textsc{mmMRNeps1e-7}$, we expect the back-reaction to be definitely negligible as $\theta_{\mathrm{d},0} \ll 1$.

Figure \ref{fig:enrichback} shows the dust ratio enrichment as a function of the initial Stokes number for $\textsc{mmMRN}$ (red) and  $\textsc{mmMRNeps1e-7}$ (pink) in the core, disk and envelope at $t_{\mathrm{core}}+2~$kyr. On top of this, the dust ratio in the disk but at $t_{\mathrm{core}}$ is plotted for $\textsc{mmMRN}$ (grey). As can be seen, the differences between the models at $t_{\mathrm{core}}+2~$kyr have similar dust-ratio enrichment. This indicates that the cumulative back-reaction of dust is negligible when $\mathrm{St}_{k,0} <  10^{-2}$ and only corrective above that. We are also confident that the mixing between dust enriched and dust depleted content, i.e. a non-negligible $\nabla \epsilon$, are the source of discrepancy between a pure exponential enrichment and the values measured in our models in the disk. We indeed observe a very good agreement with the exponential law for $\textsc{mmMRN}$ at $t_{\mathrm{core}}$. At this time the disk has just formed from the dense and dust enriched material of the inner envelope. Later at $t_{\mathrm{core}}+2~$kyr, the disk has been accreting material from the envelope that is significantly depleted in large grains which causes a diminution of the average dust enrichment.

\section{Charges on dust grains}
\label{ap:apA}
One can crudely estimate the charges on the grains during the collapse similarly to \cite{2016MNRAS.457.1037W}, to compare the relative intensities of the Lorentz and the Epstein drags. For simplicity, we assume that all the grains have the same negative charge $Z_{\mathrm{d}}$ (for all $k$, $Z_k=Z_{\mathrm{d}}$) and the ions to have a charge $Z_{\mathrm{ions}}=1$. We consider $100$ dust bins distributed as an MRN extended to millimeter-in-size grains, with a dust-to-gas ratio of $1\%$. Local electroneutrality ensures that
\begin{eqnarray}
n_{\mathrm{ions}}-n_{\mathrm{e}}+Z_{\mathrm{d}} n_{\mathrm{d}} =0,
\end{eqnarray}
$n_{\mathrm{ions}}$, $n_{e}$ and $n_{\mathrm{d}}$ being the number density of the ions, electrons and dust, the latter being given by
\begin{eqnarray}
n_{\mathrm{d}} \equiv \frac{\mu_{\mathrm{g}} m_{\mathrm{H}}}{m_{\mathrm{grain}}} \epsilon n,
\end{eqnarray}
where $m_{\mathrm{grain}}$ is the average mass of a dust grain. 

Two additional constrains are provided by the evolution equations for the ions and electrons charge density \citep{1980PASJ...32..405U,2011ApJ...743...53F}. Assuming steady state and only considering charge capture by the grains \citep{2014MNRAS.440...89K}, they give 
\begin{eqnarray}
n_{\mathrm{ions}} & = & \frac{\zeta n}{k_{\mathrm{ions, grains}} n_{\mathrm{d}}}, \nonumber \\
n_{\mathrm{e}} & =& \frac{\zeta n}{k_{\mathrm{e, grains}} n_{\mathrm{d}}},  
\end{eqnarray}
where $\zeta$ is the cosmic-ray ionization rate. Similarly to \cite{2016MNRAS.457.1037W}, we adopt the typical value $\zeta= 10^{-17} \mathrm{s}^{-1}$.  We also express $k_{\mathrm{ions, grains}}$ and $k_{\mathrm{e, grains}}$, the charge capture rates on neutral grains, as in \citep{2011ApJ...743...53F} as
\begin{eqnarray}
k_{\mathrm{ions, grains}}  \equiv \pi s_{\mathrm{grain}}^2 \sqrt{\frac{8 k_{\mathrm{b}}T_{\mathrm{g}}}{\pi m_{\mathrm{ions}}}} (1-\frac{e^2 Z_{\mathrm{d}}}{s_{\mathrm{grain}} k_{\mathrm{b}}T_{\mathrm{g}}}), \nonumber \\
k_{\mathrm{e, grains}}  \equiv \pi s_{\mathrm{grain}}^2 \sqrt{\frac{8 k_{\mathrm{b}}T_{\mathrm{g}}}{\pi m_{\mathrm{e}}}} \exp\left[\frac{e^2 Z_{\mathrm{d}}}{s_{\mathrm{grain}} k_{\mathrm{b}}T_{\mathrm{g}}}\right].
\end{eqnarray}
where $s_{\mathrm{grain}}$ is the average grain size, $m_{\mathrm{e}}$ is the electron mass and  $m_{\mathrm{ions}}$ the ions mass, assumed to be $24.3$ proton masses. Using the previous equations, one obtains
\begin{eqnarray}
Z_{\mathrm{d}}=\frac{\zeta}{n} \left(\frac{m_{\mathrm{grain}}}{\epsilon \mu_{\mathrm{g}} m_{\mathrm{H}}}\right)^2 \left[ \frac{1}{k_{\mathrm{ions, grains}} (Z_{\mathrm{d}}) }-\frac{1}{k_{\mathrm{e, grains}} (Z_{\mathrm{d}})} \right],
\end{eqnarray}
that we invert using the Newton-Raphson method to get $Z_{\mathrm{d}}$.
 
\section{Distributions}
In Table~\ref{tab:distri}, we provide the initial dust distributions, rounded quantities for the three maximum grain size used in our models (the exact calculation can be made using the method presented in \ref{sec:distri}).  Table  \ref{tab:corem} shows $\left<\Theta_{\mathrm{d},k}\right>_{m}$ (in $\%$), the dust-to-gas ratio averaged in mass, along with its corresponding dust enrichment and the gas mass (in units of $M_{\odot}$) for all runs  $t_{\mathrm{core}}+2$~kyr and all the different objects. 
\begin{table*}[t]
       \caption{Initial dust distributions, rounded quantities for the three maximum grain size used in our models (the exact calculation can be made using the method presented in \ref{sec:distri}) . For each $S_{\mathrm{max}}$, the first line represent the grain size in cm and the second the initial dust-to-gas ratio. The exponents are given by the parenthesis and the integers from 1 to 10 correspond to bins of increasing sizes.}      
\label{tab:distri}      
\centering          
\begin{tabular}{c c c c c c c c c c c c}     
\hline\hline       
 $S_{\mathrm{max}}$ & $s$ (cm), $\theta$ : & 1 & 2 & 3 & 4 & 5 & 6 & 7 & 8 & 9 & 10 \\
 \hline 
 2.5(-5) & $s$ (cm)  &  4.11(-7) &  6.08(-7)  & 8.99(-7) &   1.33(-6)
  & 1.97(-6) &  2.91(-6) &  4.3(-6) &  6.36(-6)
   &9.4(-6) &  1.39(-5) \\
   &$\theta$ &  3.56(-4)& 4.33(-4)  &5.26(-4) &6.40(-4)  &7.78(-4)  &9.46(-4)
  &1.15(-3)  & 1.4(-3)&  1.7(-3)  &2.1(-3)\\
  \hline
 0.01  &   $s$ (cm)  &  3.05(-7) &  8.2(-7) &   2.21(-6)  & 5.95(-6) & 1.6(-5) &  4.31(-5) &  1.16(-4)  & 3.12(-4)
&   8.41(-4) &  2.26(-3)  \\

 & $\theta$ & 4.56(-5)  & 7.49(-5) &   1.223(-4)  & 2.02(-4)
   &3.31(-4)   &5.43(-4)  & 8.9(-4) &  1.46(-3)
   &2.40(-3) &   3.93(-3) \\
 \hline                  
 0.1 &   $s$ (cm)  & 2.72(-7) &  9.2(-7)  & 3.12(-6) &  1.05(-5)& 3.58(-5)   &1.21(-4)  & 4.12(-4)  & 1.4(-3) &  4.73(-3) &  1.6(-2) \\
&$\theta$  & 1.88(-5) &  3.47(-5) &   6.39(-5)&   1.18(-4) &2.16(-4) &3.99(-4)  & 7.34(-4) &   1.35(-3)&  2.49(-3)  &  4.58(-3) \\
 \hline
 
\end{tabular}
\end{table*}

\begin{table*}[t!]
       \caption{Mass averaged dust-to-gas ratio $\left<{\Theta_{d,k}}\right>_{m}$ (in $\%$) and gas mass (in units of $M_{\odot}$) for all runs  $t_{\mathrm{core}}+2$~kyr and all the objects, the number inside the brackets is the corresponding mass averaged dust enrichment. When $\left<\bar{\Theta}_{\rm{d}}\right>_{m}>1$,  the enrichment is referenced in red while it is referenced in blue when $\left<\bar{\Theta}_{\rm{d}}\right>_{m}\leq1$.  We denote the different objects as follows, $\mathcal{F}_j$ represent the fragments (F0 being the FHSC), $\mathcal{D	}$ the disks, $\mathcal{O}$ the outflow, $\mathcal{P}$ the pseudo-disks and $\mathcal{E}$ the envelope. }      
\label{tab:corem}      
\centering          
\begin{tabular}{c  c c  c  c c c c}     
\hline\hline       
Model & j & $\left<{\Theta_{d,k\leq6}}\right>_{m}$ & $\left<{\Theta_{d,7}}\right>_{m}$ & $\left<{\Theta_{d,8}}\right>_{m}$ & $\left<{\Theta_{d,9}}\right>_{m}$ &$ \left<{\Theta_{d,10}}\right>_{m}$  & $ m_{\mathrm{g}}$    \\
\hline 
$\textsc{mmMRN}$ & $\mathcal{F}_0$ & 0.085[1] & 0.075[\red{1.03}] & 0.14[\red{1.03}] & 0.31[\red{1.24}] & 1.17[\red{2.6}] & 0.082 \\
& $\mathcal{F}_1$ & 0.085[1] & 0.076[\red{1.04}] & 0.15[\red{1.1}] & 0.37[\red{1.54}] & 1.38[\red{3}] & 0.024 \\
& $\mathcal{F}_2$& 0.085[1]& 0.076[\red{1.04}] & 0.15[\red{1.1]} & 0.37[\red{1.54}] & 1.37[\red{2.98}] & 0.024 \\
&$\mathcal{D}$ & 0.085[1]& 0.075[\red{1.03}] & 0.15[\red{1.1}] & 0.34[\red{1.41}] & 1.1[\red{2.4}]& 0.096   \\
&$\mathcal{E}$ &0.085[1] & 0.073[1] & 0.13[\blue{0.96}] & 0.23[\blue{0.95}] & 0.34[\blue{0.73}] & 1.0 \\
$\textsc{MRN}$ & $\mathcal{F}_0$ & 0.37[1] & 0.12[1] & 0.14[1] & 0.17[1] & 0.21[1] & 0.088 \\
& $\mathcal{F}_1$& 0.37[1] & 0.12[1] & 0.14[1]& 0.17[1] & 0.21[1] & 0.033 \\
& $\mathcal{F}_2$ & 0.37[1] & 0.12[1] & 0.14[1] & 0.17[1] & 0.21[1] & 0.033 \\
&$\mathcal{D}$  &0.37[1] & 0.12[1] & 0.14[1] & 0.17[1]& 0.21[1] & 0.091 \\
&$\mathcal{E}$ & 0.37[1] & 0.12[1] & 0.14[1] & 0.17[1] & 0.21[1] & 0.98\\ 
$\textsc{100micMRN}$ & $\mathcal{F}_0$ & 0.13[1] & 0.09[1] & 0.15[1] & 0.25[\red{1.04}] & 0.43[\red{1.1}]  & 0.086 \\
& $\mathcal{F}_1$ & 0.13[1] & 0.09[1] & 0.15[1] & 0.25[\red{1.04}]  & 0.46[\red{1.17}]  & 0.023 \\
& $\mathcal{F}_2$ & 0.13[1] & 0.09[1] & 0.15[1] & 0.25[\red{1.04}] & 0.46[\red{1.17}]  & 0.023 \\
&$\mathcal{D}$& 0.13[1] & 0.09[1] & 0.15[1] & 0.25[\red{1.04}] & 0.44[\red{1.13}] & 0.097\\
&$\mathcal{E}$& 0.13[1] & 0.09[1] & 0.15[1] & 0.24[1]  & 0.38[\blue{0.97}] & 0.99\\
$\textsc{mmMRNa0.25}$ & $\mathcal{F}_0$ & 0.085[1] & 0.074[\red{1.01}] & 0.14[\red{1.03}] & 0.26[\red{1.04}] & 0.51[\red{1.12}] & 0.20 \\
& $\mathcal{F}_1$ & 0.085[1] & 0.074[\red{1.01}] & 0.14[\red{1.03}] & 0.27[\red{1.08}] & 0.58[\red{1.29}] & 0.10 \\
& $\mathcal{F}_2$ & 0.085[1]& 0.074[\red{1.01}] & 0.14[\red{1.03}] & 0.27[\red{1.08}] & 0.58[\red{1.29}] & 0.10 \\
&$\mathcal{D}$&0.085[1] & 0.074[\red{1.01}] & 0.14[\red{1.03}] & 0.26[\red{1.04}] & 0.56[\red{1.24}] & 0.16 \\
&$\mathcal{E}$ & 0.085[1] & 0.073[1] & 0.13[\blue{0.96}] & 0.24[\blue{0.96}]& 0.4[\blue{0.89}] & 0.7 \\
\hline
 $\textsc{mmMRNmhd}$ & $\mathcal{F}_0$ & 0.085[1]& 0.076[\red{1.04}] & 0.15[\red{1.1}] & 0.37[\red{1.48}] & 1.5[\red{3.33}]& 0.075 \\
 &$\mathcal{D}$& 0.085[1] & 0.076[\red{1.04}] & 0.15[\red{1.1}] & 0.39[\red{1.56}] & 1.58[\red{3.58}] & 0.0012 \\
 &$\mathcal{P}$&0.086[\red{1.01}] & 0.077[\red{1.05}] & 0.16[\red{1.19}] & 0.43[\red{1.72~}] & 1.23[\red{2.73}] & 0.1 \\
  &$\mathcal{O}$&0.085[1] & 0.074[\red{1.01}] & 0.14[\red{1.03}] & 0.26[\red{1.04}] & 0.1[\red{0.22}] & 0.0082 \\
  &$\mathcal{E}$& 0.085[1] & 0.072[\blue{0.98}] & 0.13[\blue{0.96}] & 0.2[\blue{0.8}] & 0.17[\blue{0.38}] & 0.5 \\
 $\textsc{mmMRNnimhd}$ & $\mathcal{F}_0$ & 0.085[1] & 0.077[\red{1.5}] & 0.15[\red{1.1}] & 0.37[\red{1.48}]& 1.5[\red{3.33}] & 0.081\\
 &$\mathcal{D}$& 0.085[1]& 0.076[\red{1.04}] & 0.15[\red{1.1}] & 0.37[\red{1.48}] & 1.44[\red{3.2}] & 0.011 \\ 
 &$\mathcal{P}$ & 0.086[\red{1.01}] & 0.077[\red{1.05}] & 0.16[\red{1.19}] & 0.43[\red{1.72}] & 1.22[\red{2.71}]& 0.1\\
  &$\mathcal{O}$& 0.086[\red{1.01}] & 0.076[\red{1.04}] & 0.15[\red{1.1}] & 0.32[\red{1.28}] & 0.2[\blue{0.44}] & 0.0018\\
   &$\mathcal{E}$ & 0.085[1] & 0.072[\blue{0.98}] & 0.13[\blue{0.96}] & 0.2[\blue{0.8}] & 0.17[\blue{0.38}] & 0.5\\ 
 \hline

\end{tabular}
\end{table*}

\end{appendix}
\begin{acknowledgements}
   First, we thank the referee for providing useful comments and advice that helped us to ameliorate our manuscript. We acknowledge financial support from "Programme National de Physique Stellaire" (PNPS) of CNRS/INSU, CEA and CNES, France. This work was granted access to the HPC resources of CINES (Occigen) under the allocation DARI A0020407247 made by GENCI. Computations were also performed at the Common Computing Facility (CCF) of the LABEX Lyon Institute of Origins (ANR-10-LABX-0066). This work took part under the programs ISM3D and Core2disk of the PSI2 project funded by the IDEX Paris-Saclay, ANR-11-IDEX-0003-02.  This project was partly supported by the IDEXLyon project (contract n ANR-16-IDEX-0005) under University of Lyon auspices. The plots were generated using the very efficient Osiris library developed by Neil Vaytet, Tommaso Grassi and Matthias Gonz\'alez whom we thank. This project has received funding from the European Union's Horizon
2020 research and innovation program under the Marie
Sk\l{}odowska-Curie grant agreement No 823823. We also thank Etienne Jaupart for useful discussions on the theoretical model for the dust enrichment during the collapse.
\end{acknowledgements}

\bibliographystyle{aa}
\bibliography{main}

\end{document}